\documentclass[twocol]{ametsocV6.1}

\usepackage{amsthm,amsmath,amsfonts,amssymb}
\usepackage{graphicx,floatrow}
\usepackage{placeins}
\usepackage{booktabs}
\usepackage{verbatim}
\usepackage{bm}
\usepackage{wrapfig}
\usepackage[shortlabels]{enumitem}

\newcommand{\X}{\mathbf{X}}
\newcommand{\x}{\mathbf{x}}
\renewcommand{\S}{\mathbf{S}}

\newcommand{\Z}{\mathbf{Z}}
\newcommand{\T}{\mathbf{BT}}
\newcommand{\z}{\mathbf{z}}

\definecolor{vermilion}{rgb}{0.89, 0.26, 0.2}

\newcommand{\revtwo}[1]{{\color{black} #1}}

\title{Structural Forecasting for Short-term Tropical Cyclone Intensity Guidance}

\authors{Trey McNeely,\aff{a,d} Pavel Khokhlov, Niccol\`o Dalmasso,\aff{b} Kimberly M. Wood,\aff{c} and Ann B. Lee\aff{a}\correspondingauthor{Ann B. Lee, annlee@stat.cmu.edu}}

\affiliation{\aff{a}{Carnegie Mellon University Department of Statistics and Data Science}\\
\aff{b}{J.P. Morgan AI Research}\\
\aff{c}{Mississippi State University Department of Geosciences}\\
\aff{d}{Microsoft AI Development Acceleration Program}}

\abstract{
\revtwo{Because geostationary satellite (Geo) imagery provides a high temporal resolution window into tropical cyclone (TC) behavior, we investigate the viability of its application to short-term probabilistic forecasts of TC convective structure to subsequently predict TC intensity. Here, we present a prototype model which is trained solely on two inputs: Geo infrared imagery leading up to the synoptic time of interest and intensity estimates up to 6 hours prior to that time. To estimate future TC structure, we compute cloud-top temperature radial profiles from infrared imagery and then simulate the evolution of an ensemble of those profiles over the subsequent 12 hours by applying a Deep Autoregressive Generative Model (PixelSNAIL). To forecast TC intensities at hours 6 and 12, we input {\em operational} intensity estimates up to the current time (0 h) and simulated future radial profiles up to +12 h into a ``nowcasting'' convolutional neural network. We limit our inputs to demonstrate the viability of our approach and to enable quantification of value added by the observed and simulated future radial profiles beyond operational intensity estimates alone.  Our prototype model achieves a marginally higher error than the National Hurricane Center's official forecasts despite excluding environmental factors, such as vertical wind shear and sea surface temperature. We also demonstrate that it is possible to reasonably predict short-term evolution of TC convective structure via radial profiles from Geo infrared imagery, resulting in interpretable structural forecasts that may be valuable for TC operational guidance.}}

\begin{document}

%% Necessary!
\maketitle

%%%%%%%%%%%%%%%%%%%%%%%%%%%%%%%%%%%%%%%%%%%%%%%%%%%%%%%%%%%%%%%%%%%%%
% SIGNIFICANCE STATEMENT/CAPSULE SUMMARY
%%%%%%%%%%%%%%%%%%%%%%%%%%%%%%%%%%%%%%%%%%%%%%%%%%%%%%%%%%%%%%%%%%%%%
%
% If you are including an optional significance statement for a journal article or a required capsule summary for BAMS 
% (see www.ametsoc.org/ams/index.cfm/publications/authors/journal-and-bams-authors/formatting-and-manuscript-components for details), 
% please apply the necessary command as shown below:
%
% Significance Statement (all journals except BAMS)
%

\statement 
\revtwo{This work presents a new method of short-term probabilistic forecasting for tropical cyclone (TC) convective structure and intensity using infrared geostationary satellite observations. Our prototype model’s performance indicates that there is some value in observed and simulated future cloud-top temperature radial profiles for short-term intensity forecasting. The non-linear nature of machine learning tools can pose an interpretation challenge, but structural forecasts produced by our model can be directly evaluated and thus may offer helpful guidance to forecasters regarding short-term TC evolution. Since forecasters are time-limited in producing each advisory package despite a growing wealth of satellite observations, a tool that captures recent TC convective evolution and potential future changes may support their assessment of TC behavior in crafting their forecasts.}

%	 Enter significance statement here, no more than 120 words. See \url{www.ametsoc.org/index.cfm/ams/publications/author-information/significance-statements/} for details.
%
%% Capsule (BAMS only)
%%
%\capsule
%       Enter BAMS capsule here, no more than 30 words. See \url{www.ametsoc.org/index.cfm/ams/publications/author-information/formatting-and-manuscript-components/#capsule} for details.
%
%% * * If using twocol mode, you will need to use the commands "twocolsig" and "twocolcapsule" in place of "sig" and "capsule"
%%      to ensure that the text box correctly spans across both columns.
%

%%%%%%%%%%%%%%%%%%%%%%%%%%%%%%%%%%%%%%%%%%%%%%%%%%%%%%%%%%%%%%%%%%%%%
% MAIN BODY OF PAPER
%%%%%%%%%%%%%%%%%%%%%%%%%%%%%%%%%%%%%%%%%%%%%%%%%%%%%%%%%%%%%%%%%%%%%
%

\section{Introduction}
Tropical cyclones (TCs) are powerful, organized systems that pose a major risk to coastal populations. Though many statistical models provide forecast guidance on future TC intensity change (e.g., the Statistical Hurricane Intensity Prediction Scheme [SHIPS]; \citealt{DeMaria1999}), direct measurement of most predictors such as relative humidity or vertical wind shear used in such models is impossible due to the development of TCs over open ocean far from land-based observing networks \citep{gray1979hurricanes}. Many predictors must be inferred through a combination of remote observation and dynamic models of ocean and atmospheric behavior.

Infrared (IR; 10.3-10.7$\mu$m) imagery from geostationary (Geo) satellites such as the Geostationary Operational Environmental Satellites (GOES) provides one of the few regular high-resolution observations of TC behavior over the open ocean with a historical record spanning decades \citep{Knapp2018,mergir}. Furthermore, modern Geo IR platforms such as GOES-16 provide observations at even greater spatial and temporal resolution \citep{schmit2017closer}. Since cloud-top temperature is related to cloud-top height, low IR temperatures tend to indicate higher cloud tops and thus stronger convection, and convective structures are known to be related to TC intensity \citep{Dvorak1975,olander2007advanced}.

\begin{figure*}[!tbp]
	\centering
	\includegraphics[width=.75\linewidth]{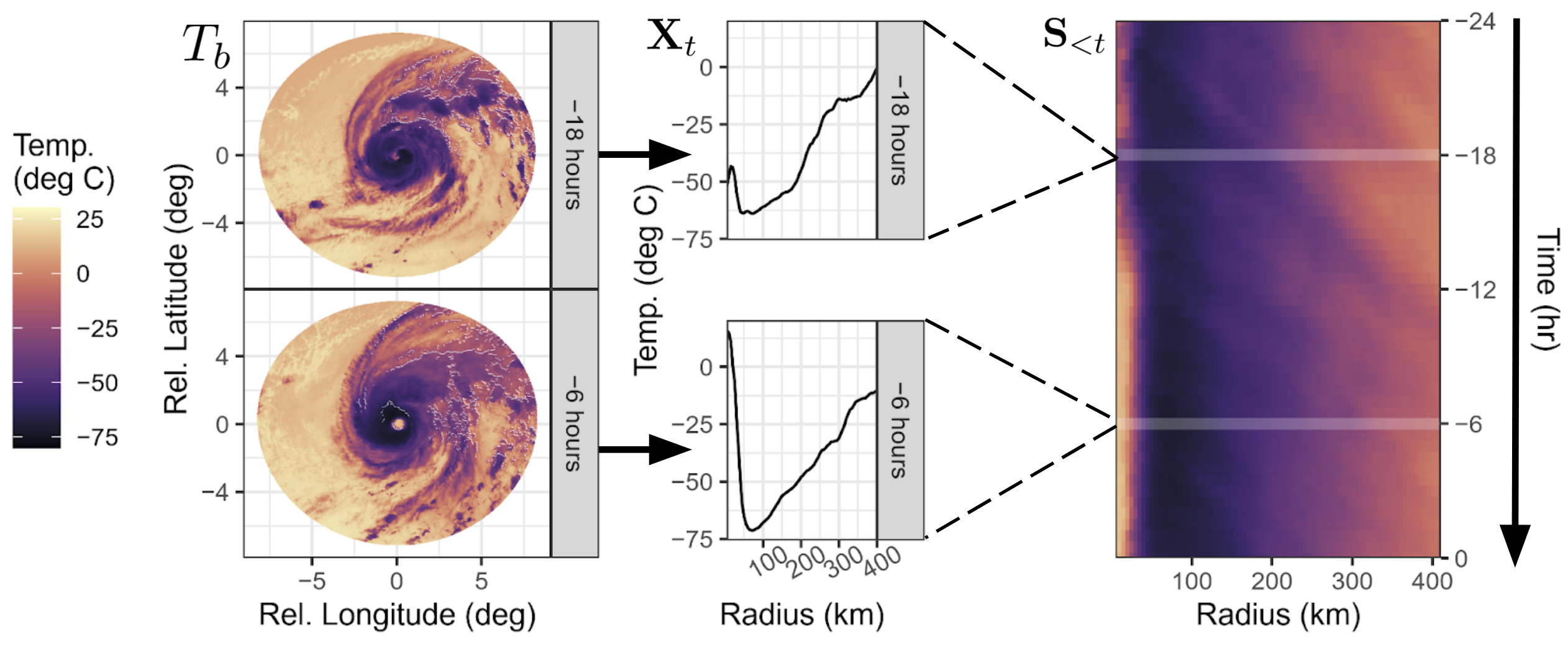}
	\caption{\textbf{Evolution of TC IR Imagery as Structural Trajectories.} The raw data at each time $t$ is a sequence of  TC-centered cloud-top temperature images from GOES. We quantify the image at time $t$ by its radial profile ($X_t$). The sequence of consecutive radial profiles, sampled every 2 hours, defines a \emph{structural trajectory} or Hovm\"oller diagram ($S_{<t}$).}
	\label{fig:profiles}
\end{figure*}

In light of this growing record of satellite observations, a broad array of recent works have explored the wealth of information contained in the \emph{spatio-temporal} structure of Geo IR imagery. The Dvorak technique and more recent Advanced Dvorak Technique (ADT) have long related Geo IR imagery to TC intensity \citep{Dvorak1975,olander2007advanced}, and more recent work has leveraged neural networks to improve the nowcasting accuracy of the ADT (AI enhanced Dvorak Technique; \citealt{olander2021investigation}). \revtwo{Here, we define ``nowcasting'' as estimating the current TC intensity based on intensity estimates up to 6 h prior and IR features up to the current time (0 h). Spatial analyses of IR imagery have been leveraged} to improve forecasts of TC eye formation, a process related to intensification \citep{demaria2015automated,knaff2017forecasting}. The deviation angle variance (DAV) technique, a measure of convective organization in IR imagery, contains valuable information for short-term ($\le$24 h) TC intensity guidance \citep{hu2020short}. The shape and evolution of Geo IR radial profiles is known to relate to intensity and intensity change respectively \citep{Sanabia2014,mcneely2020unlocking}. In this work, we utilize the {\em evolution over time} of radial profiles (see Figure \ref{fig:profiles}) to  jointly forecast short-term TC intensity and structure changes. We leverage deep auto-regressive (AR) generative models to construct {\em interpretable and high-resolution} structural probabilistic forecasts, which display entire functions rather than time series of thresholded quantities, such as pixel counts beneath a given temperature threshold.

Concurrent with the rise of high-resolution Geo IR imagery is the growing application of convolutional neural networks (CNNs), powerful tools for performing prediction tasks with images as input. Predicting TC intensity from Geo IR data is an obvious candidate application; indeed, there are dozens of such works in the machine learning literature applying CNNs to this problem, including \cite{pradhan2017tropical,combinido2018convolutional,lee2019tropical,tian2020cnn,wang2020cnn}; and \cite{zhang2021tropical}. These models achieve reasonable forecast accuracy via the traditional machine learning framework with a CNN taking IR imagery as input to directly predict intensity by, e.g.,  minimizing the average squared-error loss on independent test data. Explainable AI approaches may then use methods such as layer-wise relevance propagation, saliency  maps, and activation maps to better understand \emph{how} the model produced its point estimate \citep{mcgovern2019making,ebert2020evaluation}. For an example of explainable CNN-based TC intensity forecasting in the meteorological literature, see \cite{griffin2022predicting}.

Our proposed pipeline takes a different approach to explainability---one which remains compatible with the above tools for insight into the relationships leveraged by CNNs. Our approach (i) utilizes a dimensionality-reducing functional transformation of IR imagery prior to analysis, and (ii) provides \revtwo{12-hour ensemble forecasts of TC convective structure in addition to TC intensity}. 

First, we extract {\em scientifically-motivated functional features}, reducing the dimension of the problem (from 2-D images over time to 1-D functions over time) in a directly interpretable summary, rather than directly relying on the CNN to extract salient features from (high-dimensional and low-sample size) raw Geo IR imagery. These rich summary functions are derived from the ORB suite: Organization (e.g., DAV as a function of radius), Radial structure (e.g., the radial profiles examined in this work), and Bulk morphology (e.g., pixel counts as a function of a temperature threshold). Temporal sequences of radial profiles are highly relevant to both intensity and intensity change \citep{Sanabia2014,mcneely2020unlocking,mcneely2022detecting}. Temporal changes in these sequences of profiles \revtwo{can be visualized} via Hovm\"oller diagrams, which are more readily digestible by users than inferring temporal patterns from animations of satellite imagery.

Second, we provide a \revtwo{probabilistic} {\em structural forecast}, a prediction of \revtwo{an ensemble of} possible TC convective evolution, rather than directly predicting future intensity from past IR structure \revtwo{and TC intensity}. Our novel approach to intensity guidance via Geo IR imagery results in interpretable intensity forecasts such as ``our model predicts \revtwo{short-term intensification due to the potential} emergence of an eye-eyewall structure in the next 12 hours''. Though methods such as layer-wise relevance propagation can provide further insight into the CNN's use of structural forecasts, the IR structural forecasts themselves are the core of our proposed intensity guidance pipeline.

\begin{figure*}
	\centering
	\includegraphics[width=.75\textwidth]{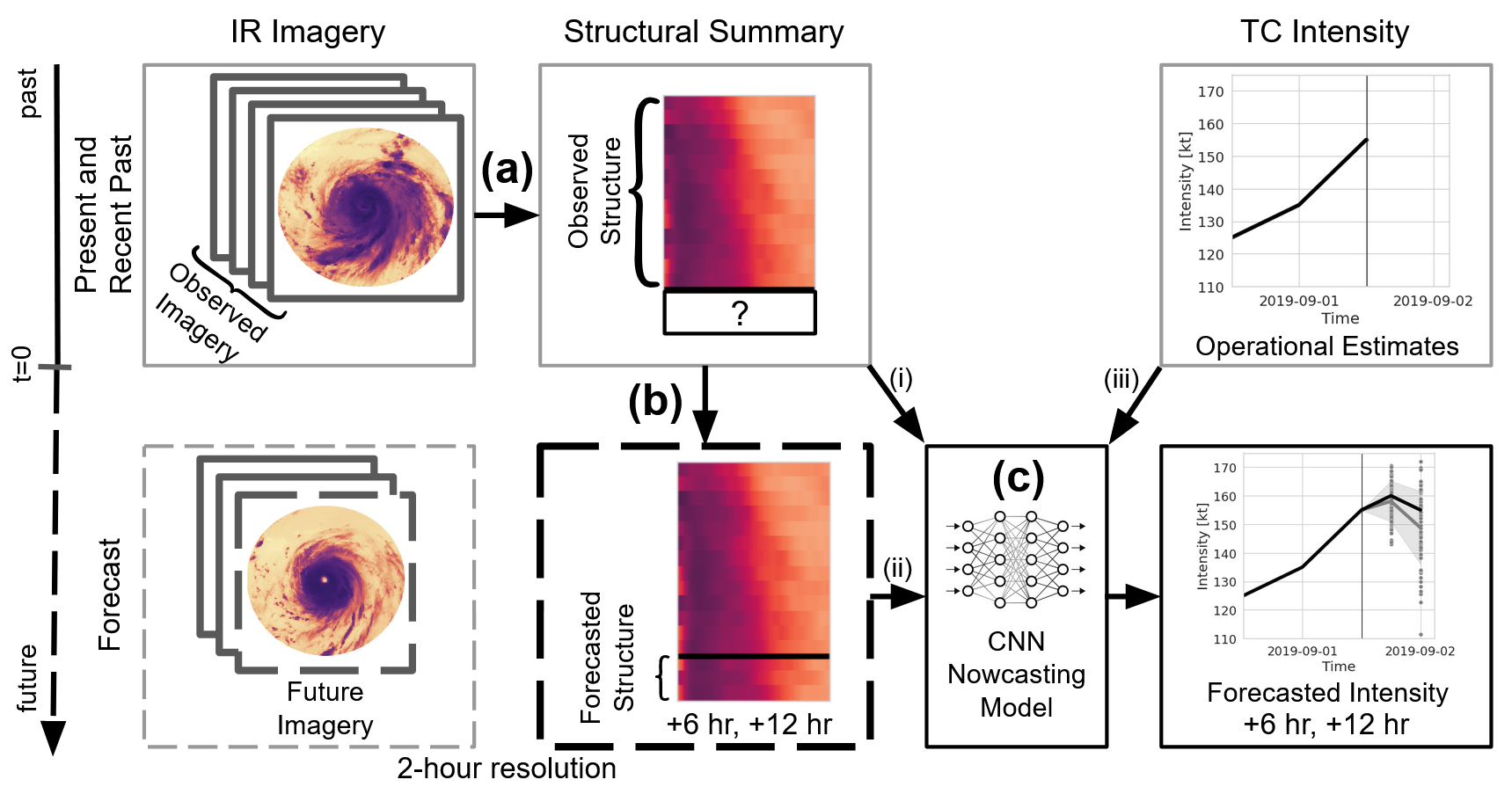}
	\caption{\textbf{TC Intensity Guidance via Structural Forecasting.} Outline of the structural forecasting to intensity guidance pipeline. {\bf(a)} ORB functions are used to quantify the evolution of spatio-temporal convective structure, linking IR imagery to Structural Summaries. {\bf(b)} We generate structural forecasts by projecting the ORB functions into the future via a deep autoregressive model, thereby filling in the missing +6 hour and +12 hour structure.
		{\bf(c)} A CNN nowcasting model then forecasts intensities at +6 to +12 hours from three sources of inputs: (i) observed structure, (ii) forecasted structure, and  (iii) \revtwo{operational intensity estimates up through the current time}.}
	\label{fig:structural_forecasts}
\end{figure*}

Figure \ref{fig:structural_forecasts} outlines Section~\ref{sec:methods} via a schematic diagram of the structural forecasting to intensity forecasting pipeline. There are three main subsections:
\begin{enumerate}[(a)]
\item {\em Structural trajectories via ORB.} First, we apply the ORB framework \citep{mcneelyquantifying,mcneely2020unlocking} to observed IR imagery to create a ``structural summary'' (Figure \ref{fig:profiles}) of the spatio-temporal evolution of the present and recent past TC structure.

\item {\em Structural forecasting \revtwo{with a} deep autoregressive generative model.} Next, we propagate the observed IR structure up to 12 hours forward in time via a deep pixel-autoregressive model, which stochastically simulates \revtwo{an ensemble of} possible trajectories of IR radial profiles.

\item {\em \revtwo{Forecasting} TC intensity via convolutional neural networks.} Finally, we input the observed structure, the forecasted structure, and \revtwo{TC intensity up to 6 hours prior to the current time into a nowcasting model to estimate the current intensity}; we choose CNNs because they are easy to train and commonly used for image data. \revtwo{By filling in the missing $t+6$ hour and $t+12$ hour structure, we can then extend the nowcasting model from a nowcast for time $t$ (i.e., hour 0) to a forecast at time $t+6$ hours and then to time $t+12$ hours.}
\end{enumerate}

Section~\ref{sec:model_results} details the results of our prototype forecasting pipeline. The final Geo IR-based TC intensity guidance provides inherent measures of uncertainty and \revtwo{insight into the potential TC structural changes that influence} a given forecast. The results in this work use proof-of-concept structural forecasting and \revtwo{a pipeline that relies} solely on persistence predictors (i.e., prior intensity \revtwo{estimates}) together with \revtwo{observed past and simulated future} radial profiles; no environmental factors such as vertical wind shear or ocean heat content are included at this time. We demonstrate that a purely autoregressive prototype achieves a useful degree of forecasting accuracy. 

\section{Data}
\revtwo{Our model relies on two data sources: sequences of Geo IR imagery captured by GOES satellites and past TC intensity. For training and verification (that is, model selection), we use NOAA's Hurricane Database 2 (HURDAT2; Landsea and Franklin 2013) because that database provides the post-season best estimates of TC intensities. For forecasting, we rely on \emph{operational} TC intensity estimates, the CARQ entries from the Naval Research Laboratory’s Automated Tropical Cyclone Forecast (ATCF) operational “A-deck” files (Sampson and Schrader 2000) to assess model performance under real-time conditions.}

GOES IR imagery is available through NOAA's Merged IR (MERGIR) database \citep{mergir} at 30-minute$\times$4-km resolution over the North Atlantic (NAL) basin from 2000--2020. For each TC, we download $\sim$2,000 km$\times$2,000 km ``stamps'' of IR imagery centered on the TC location at a 30-minute temporal resolution. Figure \ref{fig:profiles} (left) shows two such stamps after an 800-km radius mask is applied. For this work, we sample the 30-minute data at 2-hour resolution because of periodic corruption of the imagery in the MERGIR database \citep{zhongemail}.

During training, we linearly interpolate TC location and intensity from HURDAT2 to obtain locations and intensities for non-synoptic times; however, model assessment is restricted to synoptic times. We include TC lifetimes between the first synoptic time at which intensity reaches at least 35 kt and the last synoptic time at which intensity is at least 35 kt; note that this can result in the inclusion of 
TCs $<$ 35 kt if the TC decays and then re-intensifies. 

\revtwo{Finally, we rely on NHC's official forecast verification to assess our model's performance. We also} draw on the SHIPS developmental database's 200-850-hPa vertical wind shear values calculated within a 200-800-km annulus from the TC center  \revtwo{as reference} during model validation due to the \revtwo{known} impact of shear on TC convective structure \citep{SHIPSdev}.

\section{Methods}\label{sec:methods}
As outlined in Figure \ref{fig:structural_forecasts}, we first construct a summary of IR structural evolution ({\bf{a}}; Section \ref{sec:methods}\ref{sec:orb}). We then train a stochastic autoregressive model, which is an explicit likelihood model (of structural trajectories) that we can use to simulate probable IR structural evolution ({\bf{b}}; Section \ref{sec:methods}\ref{sec:pixel}). Finally, we combine observed and forecasted structure with  \revtwo{operational intensity estimates up to and including the current time} to provide interpretable short-term intensity guidance, based solely on Geo IR imagery and operational intensity estimates ({\bf{c}}; Section \ref{sec:methods}\ref{sec:CNN}).

\subsection{Structural Trajectories via \texttt{ORB}}\label{sec:orb} 
Operational forecasting of TC intensity is a human-in-the-loop process and thus places a premium on guidance interpretability. In this spirit, the ORB framework (Organization, Radial structure, Bulk morphology) summarizes 2-D imagery via continuous 1-D functions to enable static visualization of spatio-temporal patterns in TC development via Hovm\"oller diagrams \citep{hovmoller1949trough}. Our past work focused on the rich quantification of spatial information in Geo IR imagery \citep{mcneelyquantifying,mcneely2020unlocking}. More recently, we demonstrated the value of \emph{temporal} patterns in ORB functions \citep{mcneely2022detecting}, specifically the radial profile.

The radial profile of brightness temperature $\overline{BT}(r)=\frac{1}{2\pi}\int_0^{2\pi}T_b(r,\theta)d\theta$ captures the structure of cloud-top brightness temperature ($T_b$) as a function of radius $r$ from the TC center and serves as an easily interpretable description of the depth and location of convection near the TC core \citep{Sanabia2014,mcneely2020unlocking}. The radial profiles are computed at 5-km resolution from 0-400 km ($d=80$) (Figure \ref{fig:profiles}, center); {we denote the summary of convective structure at each time $t$ by $\T_t$. The \emph{structural trajectory} is then defined as the 24-hour sequence of present and 12 preceding radial profiles at a 2-hour resolution:
 \begin{equation}\label{eq:trajectory} \revtwo{\S_{<t}=(\T_{t-24h}, \T_{t-22h},\dots,\T_{t})} \end{equation}
We visualize such a trajectory with a Hovm\"oller diagram (see Figure \ref{fig:profiles}, right).

\cite{mcneely2022detecting} demonstrated a relationship between TC intensity change and Hovm\"oller diagrams of radial profiles. However, the radial profile, if averaged over all angles, will disregard asymmetry within the original 2-D images, \revtwo{which can degrade performance for cases affected by strong vertical wind shear}. In this work, we instead compute a separate radial profile for each geographic quadrant (NE, NW, SE, SW) to capture asymmetries via the differences between quadrants. We use geographic quadrants instead of motion-relative or shear-relative quadrants because the directions of motion and shear are unstable when the magnitudes of those vectors are small.

\subsection{Structural Forecasting via Deep Autoregressive Generative Model}\label{sec:pixel}
The crucial step in our guidance framework is the propagation of radial profiles into the near future. The Hovm\"oller diagram captures the spatio-temporal evolution of the TC over an extended period of time; that is, we can summarize TC development by an easily-interpretable image. By treating the structural trajectory as an image, where the y-axis corresponds to the passage of time, forecasting radial profiles becomes equivalent to an image completion problem. That is, we predict the missing pixels at the bottom of an image (forecasted structure) given those at the top (observed structure). Image completion is an active research area in machine learning; here we focus on a state-of-the-art model in the class of pixel-autoregressive models \citep{van2016pixel}.

Pixel-autoregressive models impose an ordering on the pixels of an image, such as a raster-scan ordering (left-to-right, top-to-bottom). Let each pixel in a four-quadrant radial profile trajectory be represented by $\x_i:=(x_{i1},x_{i2}, x_{i3}, x_{i4})$, where $i$ is the index in the raster scan. The pixel-AR approach factors the joint distribution of pixel values in the image as a product of conditionals,
\begin{equation}\label{eq:AR}
    p(\x_1,...,\x_n)=\prod_{i=1}^np(\x_i|\x_{i-1},...,\x_{1}),
\end{equation}
where the probability of each pixel value is conditioned on all previous pixels in the raster scan.
Then, to generate a new radial profile, one simulates repeatedly from $p(\x_i|\x_{i-1},...,\x_1)$; due to the raster-scan ordering, the distribution of a given pixel is not influenced by elements further down the sequence, hence enforcing causality.

The challenge of how to estimate the conditional likelihoods $p(\x_i|\x_{i-1},...,\x_1)$ has given rise to many flavors of pixel-autoregressive models, including PixelRNN \citep{van2016pixel}, PixelCNN \citep{van2016conditional}, PixelCNN++ \citep{salimans2017pixelcnn++}, and PixelSNAIL \citep{chen2018pixelsnail}. This work utilizes the last model, PixelSNAIL. There are two main ingredients in the model: (i) causal convolution and (ii) self-attention. Causal convolution utilizes the same convolutional feature extraction outlined in Section \ref{sec:CNN} but masks each convolution so that each element in the raster sequence only receives information from previously generated sequences (e.g., Figure \ref{fig:mask}). Purely convolutional models, however, are restricted to small neighborhoods of pixels, leading to only a finite receptive field (area of the source image involved in a given convolution), and thus struggle with long-range dependencies in the conditional $p(\x_i|\x_{i-1},...,\x_1)$. PixelSNAIL, on the other hand, features a self-attention mechanism that leads to unbounded receptive fields with pinpointed access to information far away in the sequence; see \cite{chen2018pixelsnail} for details on the PixelSNAIL architecture.
%FIGURE

To ensure that cloud-top temperatures remain bounded, we rescale the data to the range ${\X}\in(0,1)^4$ and work in the logit-transformed space, $\Z=\log({\X}/(1-{\X}))$; while values of $\Z$ are unbounded, ${\X}$ remains bounded in $(0,1)^4$ and are then transformed back to the temperature range observed in the training data. We model the density $p(\z_i|\z_{i-1},...,\z_1)$ as 4 independent mixtures of logistic distributions, one for each quadrant. That is, for pixel $i$ in quadrant $q$, 
\begin{align}
    p(z_{iq}|\z_{i-1},\dots,\z_1)=&\sum_{k=1}^K\pi_{qk}f(z_{iq};\mu_{qk},s_{qk}) \nonumber\\
    \text{where }f(z_{iq};\mu_{qk},s_{qk})=&\frac{g_{qk}(z)}{s_{qk}(1-g_{qk}(z))^2}, \nonumber\\
    g_{qk}(z)=&\exp{\big(-(z-\mu_{qk})/s_{qk}\big)},  \nonumber\\
    \text{and }\sum_{k=1}^K\pi_{qk}=&1.  \nonumber 
\end{align}
Thus, with four quadrants and $K$ mixture components, the distribution $p(\z_i|\z_{i-1},...,\z_1)$ has $4(3K-1)$ parameters: a mean $\mu$, a scale $s$, and a mixture coefficient $\pi$ for each quadrant, with the constraint that the mixture coefficients in each quadrant sum to one. With $K=3$ mixture components, this results in 32 parameters total. Each draw from the distribution is transformed back to the bounded space via the relationship $\X=\frac{1}{1+\exp{(-\Z)}}$, then rescaled to the range of values observed in the input radial profiles.

This autoregressive model enables stochastic simulation of structural trajectories based on Geo IR persistence. For a given synoptic time, we can simulate many trajectories from the observed history and then feed each potential trajectory through the nowcasting model to obtain the associated intensity guidance. Via multiple simulations per forecast time, an ensemble forecast provides a measure of uncertainty in both structural trajectories and intensities while also offering insight into cases where the model over- or under-estimates intensity. For example, overestimates may be caused by too-low profile temperatures or overestimated symmetry between quadrants.

\begin{figure}%[!bt]
  \centering
  \includegraphics[width=.65\linewidth]{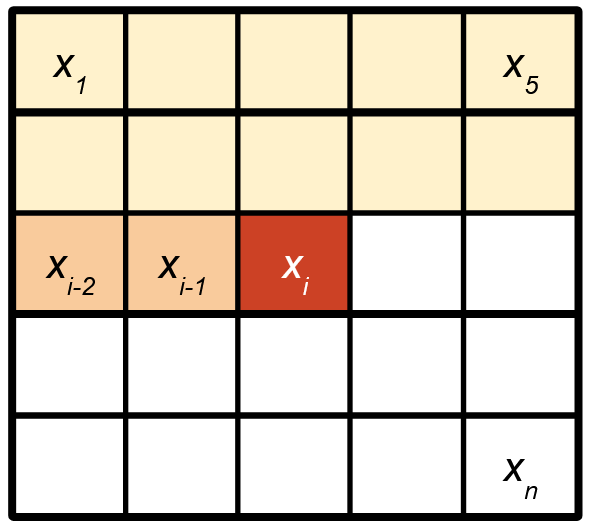}  
  \caption{\textbf{Masking in Pixel Autoregression.} Illustration of raster-scan ordering and the causal masking. Convolutions at index $i$ only have access to pixel values in previous rows (earlier time points, color coded by yellow), and pixel values in the same row but to the left of pixel $x_i$ (same time point, color coded by orange).}
  \label{fig:mask}
\end{figure}

The structural forecasting model is trained on TCs from 2000-2012, with 2013-2020 withheld for testing. We train the model using input radial profiles calculated every 2 hours but test on synoptic times. Because AR models are likelihood-based, we can directly calculate and minimize the negative log-likelihood (NLL), a measure of the model's ability to generalize well on withheld data.

\subsection{Nowcasting TC Intensity via Convolutional Neural Networks}\label{sec:CNN}
Traditional linear models are attractive for reasons of interpretability and good performance in low sample size settings. However, linear models often struggle to capture the complex, time-varying  processes which drive TCs. It is also unclear how to include the radial profile Hovm\"oller diagrams as inputs to a linear model without sacrificing interpretability. In this work, we instead consider a simple convolutional neural network to map observed IR trajectories ($\S_t$) to current intensities ($Y_t$). Because we treat time as a spatial dimension in these diagrams and a structural trajectory is represented as an image, a CNN will leverage both spatial and temporal patterns in the data.

CNNs operate by two main elements: convolutional layers and fully-connected layers. The convolutional layers first convolve each layer (here, each quadrant) with a library of filters (i.e., matrices whose entries are learned parameters); some of these filters may resemble familiar matrices, such as gradient approximators (e.g., Sobel matrices). After each convolutional layer, the image is pooled to reduce the image size and increase the receptive field of the next set of convolutions. In the final step, the results of all convolutions are passed into a fully-connected layer which approximates the relationship between the convolutional feature map and the response.

\begin{figure*}
    \centering
\includegraphics[width=.8\linewidth]{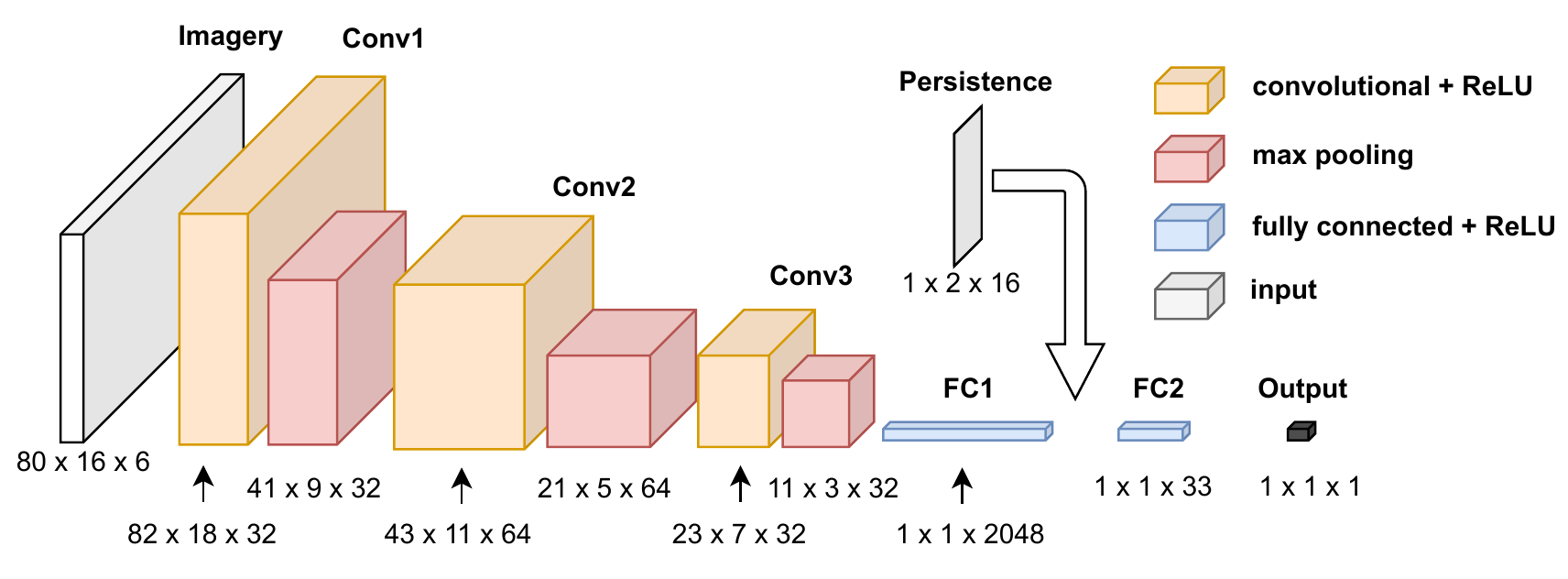}
    \caption{\textbf{Nowcasting Model Architecture.} The convolutional neural network used for nowcasting consists of three convolution-max pool layer pairs (Conv\#), fully connected layers (FC\#), and a concatenation with persistence features between regressions.}
    \label{fig:cnn_architecture}
\end{figure*}

We augment the traditional CNN in two ways. First, we add two layers which encode the radial and temporal location of pixels within the image; regular CNNs are translation invariant, whereas patterns in TCs have different meaning depending on their location and time of occurrence (corresponding to column versus row index, respectively, in the Hovm\"oller diagram). Second, we augment the model output with relevant TC persistence features (intensities and intensity changes up to 6 hours prior to time $t$)
before passing them to a second fully-connected layer. The final model is then 
\begin{align}
    Y_t=f_\text{nwcst}&(\S_{<t}, Y_{t-30h},\nonumber\\
    &Y_{t-24h}, \ldots,Y_{t-6h},\Delta Y_{t-30h},\Delta Y_{t-24h}, \ldots,Y_{t-6h})+\epsilon_t,
\end{align}
\revtwo{where $\S_{<t}$ is the 24-h structural trajectory up to current time (see Equation~\ref{eq:trajectory}); the $Y$'s are intensity data at 6-h resolution; the $\Delta Y$'s are intensity changes interpolated (ip) from 6-h data to a 2-h resolution (e.g., $\Delta Y_{t-30h}=Y_{t-30h}-Y^{\rm ip}_{t-32h}$), and $\epsilon$ is the prediction error.}

Like the structural forecasting model, the nowcasting model is trained on TCs from 2000-2012, here by minimizing the mean squared error. The model is trained on data with a 2-hour resolution (rather than synoptic times alone) with intensities linearly interpolated to those times; we do not include non-synoptic times in the test TCs (2013-2020). The details of the CNN architecture are given in Figure \ref{fig:cnn_architecture}.

\begin{figure*}[t]
	\centering
	\includegraphics[width=.8\textwidth]{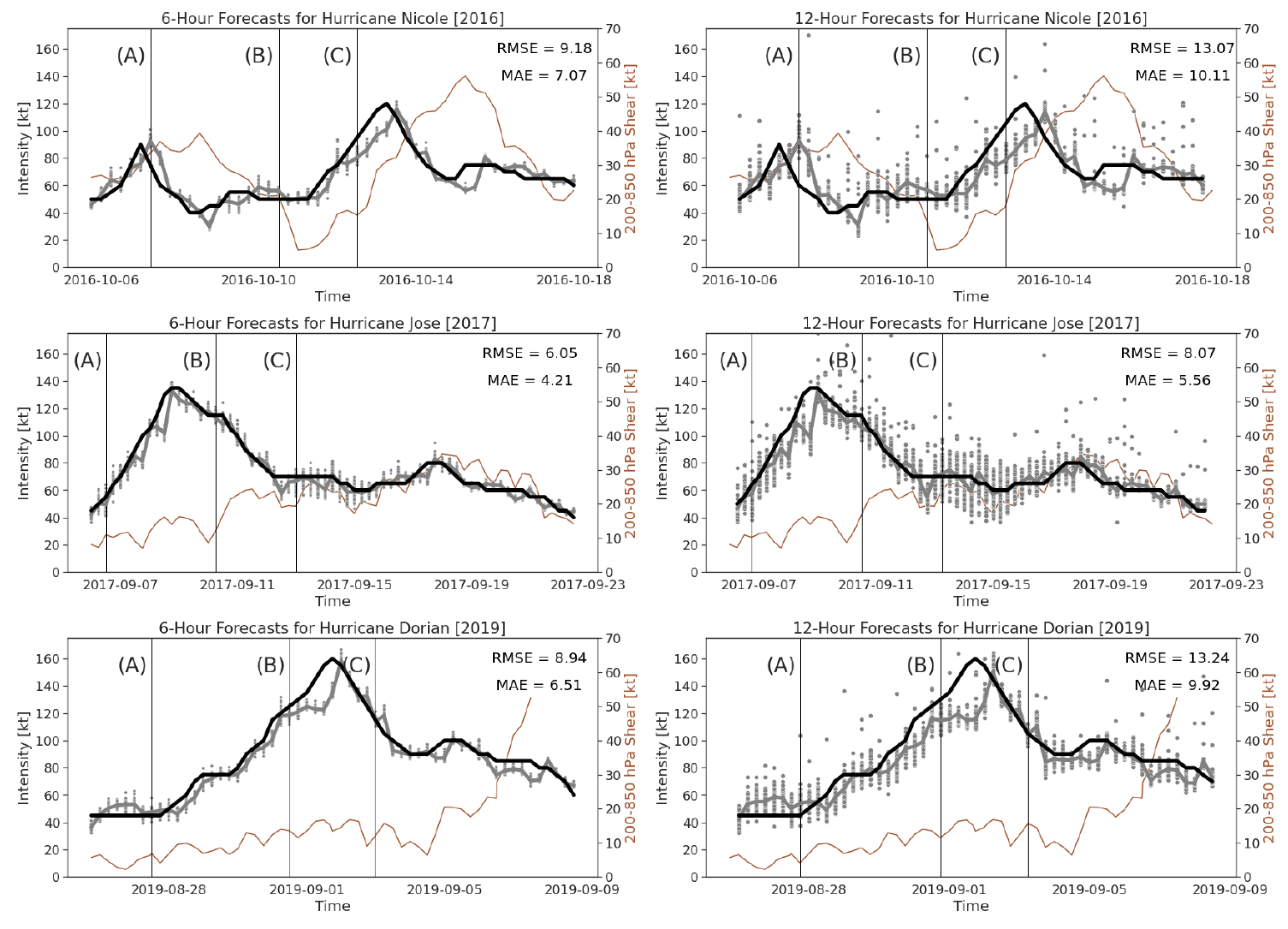}
	\caption{\textbf{Case Studies:} Comparison of the 6-hour (left) and 12-hour (right) intensity forecasts (gray) from the structural forecasting model versus observed intensities (black). Vertical wind shear also shown (gold). The solid gray lines indicate the mean of 64 simulations at each time point. Reported errors indicate the error of the average model prediction, not individual simulations. The model captures the behavior of Hurricanes Jose [2017; center] and Dorian [2019; bottom] fairly well, but struggles with both rapid weakening events exhibited by Hurricane Nicole [2016; top] as well as low-intensity maintenance periods in Hurricanes Nicole and Dorian.}
	\label{fig:case_studies}
\end{figure*}

\subsection{From Nowcasting to Forecasting} 
Section \ref{sec:methods}\ref{sec:CNN} defines a nowcasting model for \revtwo{estimating} intensity at time $t$ (i.e., hour 0)  \revtwo{by training on post-season (best-track or HURDAT2) intensities from -30 h to -6 h and imagery from -30 h to 0 h. After we have trained and validated the nowcasting model to estimate 0-h intensities, we apply the CNN nowcasting model to TC intensity forecasting}.  To \revtwo{forecast} intensity at time $t+6$ h, we need the intensities at times $\le t$  \revtwo{(in this work, we use operational intensities drawn from CARQ in the A-deck files when generating TC intensity \emph{forecasts})} 
and structural trajectories at times $\le t+6$ h (observed at times $\le t$ \revtwo{and simulated at times from $t+2$ h to $t+6$ h}).  \revtwo{Using} the structural forecasting model in Section \ref{sec:methods}\ref{sec:pixel}, we simulate many possible trajectories from times $t+2$ hr to $t+6$ hr. Each of these possible future trajectories is then passed to the nowcasting model to obtain a separate intensity forecast, 
\revtwo{giving an ensemble of possible intensities}.

Our proposed framework for intensity forecasts  \revtwo{at +6 and +12 h} has two primary benefits: (i) by providing an additional {\em structural}  forecast, we provide insight into potential TC evolution predicted by the model, such as deepening convection or the emergence of an eye; (ii) because the structural forecast is stochastic, we can straightforwardly assess the uncertainty in structural evolution over time and the associated uncertainty in the intensity forecasts.

\section{Model Results}\label{sec:model_results}
We first demonstrate the performance of our proposed model on specific cases (Hurricanes Jose [2017], Nicole [2016], and Dorian [2019]) in Section \ref{sec:case_studies}, discussing both accuracy and the insight provided by structural forecasts at 6- and 12-hour lead times. We then assess the performance of the model during 2013-2020 in the North Atlantic basin at 6- and 12-hour lead times in Section \ref{sec:validation}.

\begin{figure*}
	\centering
	\includegraphics[width=\textwidth]{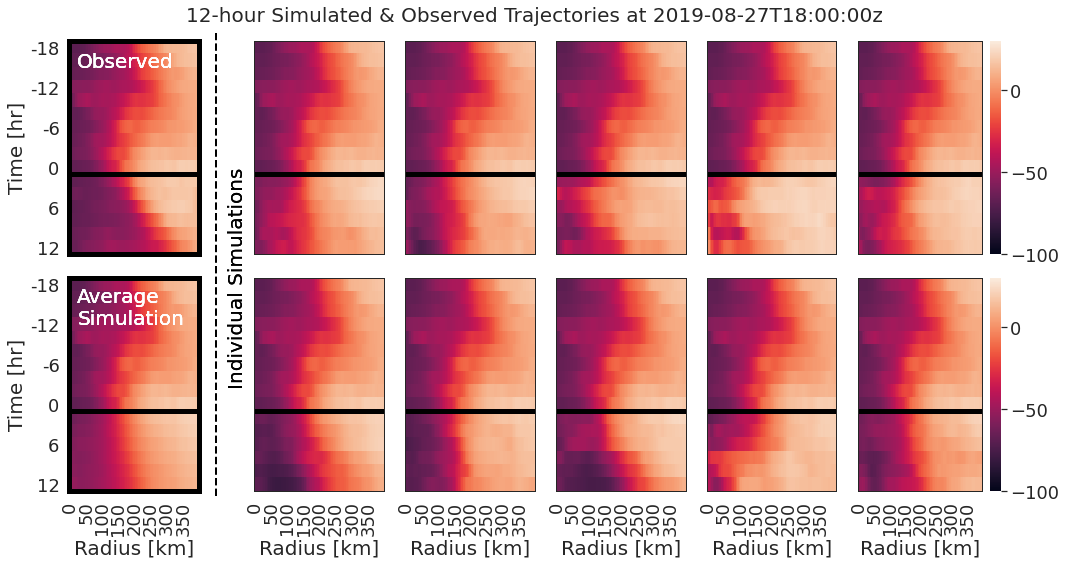}
	\caption{\textbf{Observed and Simulated Trajectories, Hurricane Dorian [2019]:} The {\em observed} structural trajectory is shown in the top left corner. To the right, we see 10 {\em individual simulations} of radial profiles (averaged over all quadrants) at 12-hour lead times. All radial profiles above the black horizontal line are observed, while profiles below the black line are simulated. The bottom left corner shows the {\em arithmetic mean} over 64 simulated trajectories.}
	\label{fig:Dorian_guessing_game}
\end{figure*}

\subsection{Case Studies}\label{sec:case_studies}
We examine Hurricane Jose (2017) due to the presence of high vertical wind shear which produces convective asymmetries not captured by the azimuthally-averaged radial profiles (i.e., not quadrant-based) of \cite{mcneely2022detecting}. Hurricane Nicole (2016) was selected due to undergoing two rapid intensification and two rapid weakening events. Finally, Hurricane Dorian (2019) was a powerful TC with many \emph{in situ} observations.

\paragraph{Intensities} 
Figure \ref{fig:case_studies} shows the 6-hour forecasts based on 64 independently simulated structural trajectories per synoptic time. Because the structural forecasts are currently based entirely on persistence---no environmental fields, such as 200-850-hPa vertical wind shear, have been included---we expect the guidance to be most useful in the short term (6- to 12-hour time frame). The steadier development of Hurricanes Jose and Dorian are well-modeled, but the swift intensity changes exhibited by Hurricane Nicole as well as the rapid intensification period of Hurricane Dorian both prove challenging to capture.

\begin{figure*}[t]
	\centering
	\includegraphics[width=\textwidth]{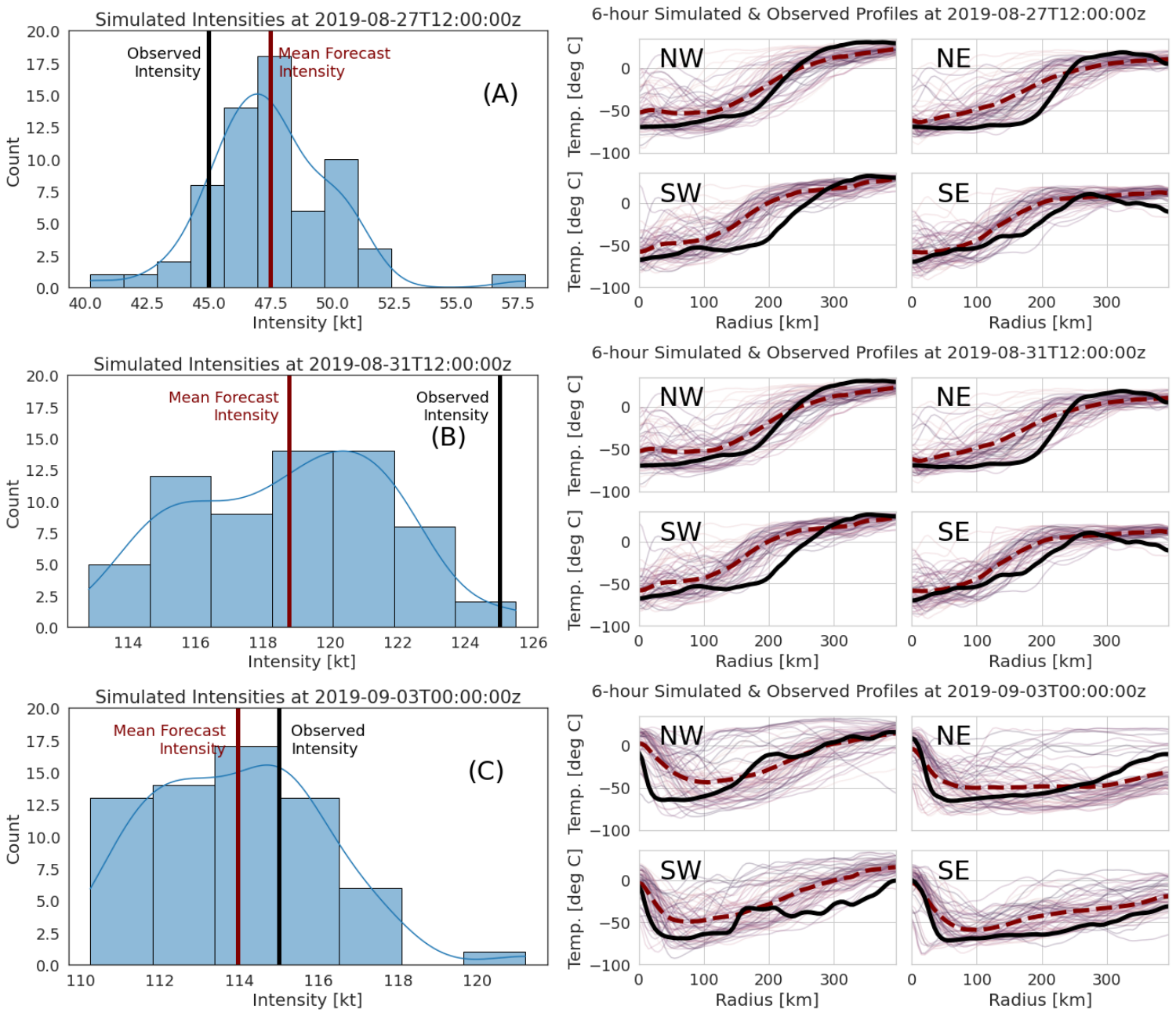}
	\caption{\textbf{Hurricane Dorian [2019] 6-h Guidance:} (Left) Distribution of forecasted intensities with observed (black) and average forecast (red) intensities marked. 
		\revtwo{The distribution of the 64 intensity forecasts in the ensemble is approximated by a histogram (bar plot) and by a kernel density estimate (blue curve).} 
		(Right) Simulated profiles by quadrant with observed profiles represented by solid black curves, and  averaged simulated profiles represented by dashed red curves.}
	\label{fig:dorian_single_time_guidance}
\end{figure*}

\begin{figure*}[t]
	\centering
	\includegraphics[width=\textwidth]{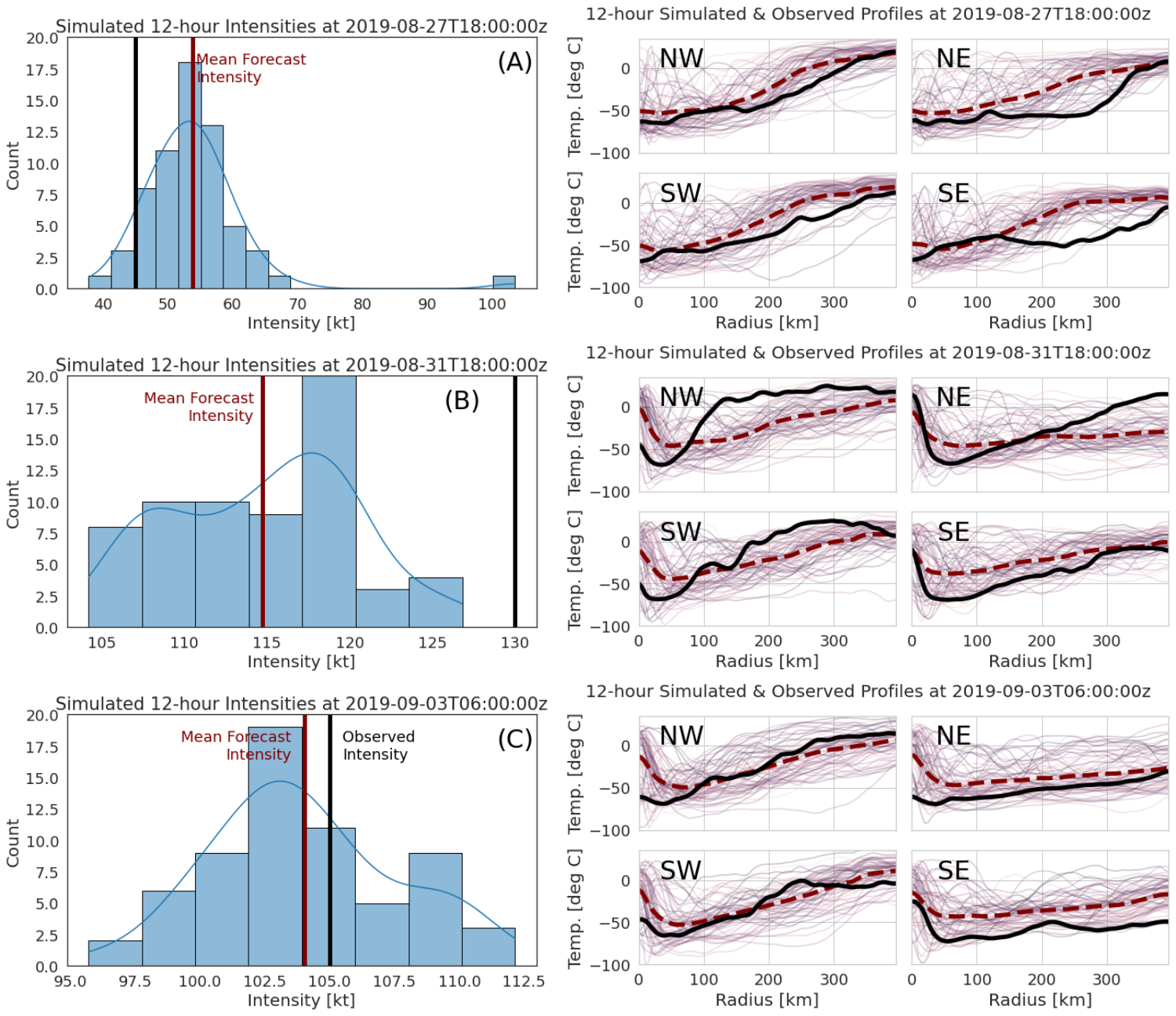}
	\caption{\textbf{Hurricane Dorian [2019] 12-h Guidance:} Model bias during intensification (center row) is more exaggerated at 12-hour lead times. (Left) Distribution of forecasted intensities with observed (black) and average forecast (red) intensities marked. \revtwo{The distribution of the 64 intensity forecasts in the ensemble is approximated by a histogram (bar plot) and by a kernel density estimate (blue curve).} (Right) Simulated profiles by quadrant with observed profiles represented by solid black curves, and  averaged simulated profiles represented by dashed red curves.}
	\label{fig:dorian_single_time_guidance_12hr}
\end{figure*}

Extending lead time to 12 hours increases the variation among individual simulations, but the average simulated intensity continues to roughly track the observed intensities. The rapid intensity change events exhibited by Hurricane Nicole are challenging to forecast with only IR persistence. However, the model follows Hurricane Jose's evolution relatively well, indicating that the model has value as-is at 12-hour lead times.

\paragraph{Diagnostics} While the end goal of intensity guidance models is ultimately prediction of TC intensity, \revtwo{our} structural forecasting pipeline  \revtwo{adds} valuable diagnostic insight into \revtwo{structural} factors contributing to its predictions. 

Figure \ref{fig:Dorian_guessing_game} demonstrates the three-step (12-hour lead time) structural forecast for Hurricane Dorian \revtwo{valid for} 18 UTC 27 August during a period in which it maintained 45-kt intensity. The final 6 rows of each Hovm\"oller diagram are simulated from the structural forecast model; in this figure, we average the four quadrants for ease of visualization (see Appendix A, Figure B10, in the supplementary materials for Hurricane Dorian structural forecasts broken down by quadrant). Cloud-top temperature magnitude tends to be underestimated, but the expansion of cloud coverage \revtwo{during this 12-h period} is captured across most simulations.

Figure \ref{fig:dorian_single_time_guidance} demonstrates the 6-hour forecasting guidance available at individual synoptic times. The average simulated profiles in each quadrant tend to track observed profiles reasonably well, although they \revtwo{tend to} predict too flat a curve and too symmetric an eye. Figure \ref{fig:dorian_single_time_guidance_12hr} \revtwo{shows the same information but for}  the 12-hour lead time. Here, model biases tend to be amplified by longer lead times. We note that the emergence of an eye \textit{is} captured in trajectory (B), even 12 hours out.

Similar figures for Hurricanes Jose and Nicole are provided in the supplemental material. In general, the structural forecast follows the observed profile, even at 12-hour lead times. We did not perform any data augmentation during training (e.g,. rotation) in order to preserve dominant geographic patterns (e.g., the prevalence of TC convection sheared eastward and northeastward in the North Atlantic), but it is possible that augmentation by rotating TCs would improve simulation fidelity, as it has been shown to improve accuracy in other TC intensity forecasting applications such as \cite{griffin2022predicting}.

\subsection{Model \revtwo{Verification}}\label{sec:validation}
The same models are used to produce 16 simulated trajectories with associated intensity guidance for each synoptic time from 2013-2020 at the 6- and 12-hour lead times. (We use 16 rather than 64 simulations when validating over the entire 8-year period for computational reasons.) Intensity predictions provided via averaging the 16 simulations are validated against HURDAT2 best-track intensities, and  \revtwo{past TC intensity values} provided as input to the model come from operational estimates (CARQ) to emulate real-time performance.

Table \ref{table:overall} (left) reports \revtwo{the performance of the simulated trajectories versus forecast lead time in terms of} root mean variance (RMV), mean absolute deviation (MAD), and bias averaged over all quadrants and radii. Let $\overline{BT}_{qi}(r)$ denote the $i^\text{th}$ simulated profile in quadrant $q$ and $\tau_q(r)$ denote the true profile in quadrant $q$. Then, 
\begin{align}
   \mathrm{RMV} = &\Big(\frac{1}{400\mathrm{\ km}}\int_0^{400}\frac{1}{4n}\sum_{q=1}^4\sum_{i=1}^n(\overline{BT}_{qi}(r)-\tau_q(r))^2dr\Big)^\frac{1}{2},\\
    \mathrm{MAD} = &\frac{1}{400\mathrm{\ km}}\int_0^{400}\frac{1}{4n}\sum_{q=1}^4\sum_{i=1}^n\big|\overline{BT}_{qi}(r)-\tau_q(r)\big|dr,\\
    \mathrm{Bias} = &\frac{1}{400\mathrm{\ km}}\int_0^{400}\frac{1}{4n}\sum_{q=1}^4\sum_{i=1}^n\overline{BT}_{qi}(r)-\tau_q(r)dr.
\end{align}
The above are defined for simulations at a single simulation time; to combine over multiple simulation times, we average MAD and bias, and average RMV in quadrature. The measures of noise (RMV and MAD) are large even for the shortest forecast lead times; increasing the simulation size beyond 16 will reduce the impact of this noise on the average forecast.  Bias, meanwhile, becomes steadily more negative with time \revtwo{for our structural forecasts (top left)}. This will lead to overestimates of intensity, as low IR temperatures are generally associated with stronger TCs. \revtwo{Persistence IR forecasts (bottom left) offer
a less biased IR forecast on average but higher overall errors in structure at all lead times.} 

Table \ref{table:overall} (right) reports \revtwo{verification statistics} for intensity guidance using the traditional definitions for root mean squared error (RMSE), mean absolute error (MAE), and bias. As expected, the negative bias in structural forecasts manifests as a positive bias in intensity guidance. 

\begin{table*}[p!]
	\begin{minipage}{.66\textwidth}
		\centering
		{Trajectory \revtwo{Verification}}\\
		\begin{tabular}{lr|rrrrrr}
			\toprule
			Model & Lead time &    2 h &    4 h &    6 h &    8 h &   10 h &   12 h \\
			\midrule
			Structural & RMV     & 10.5C & 14.7C & 17.4C & 22.2C & 25.1C & 26.9C \\
			& MAD     &  7.1C & 10.5C & 12.8C & 15.6C & 17.6C & 19.0C \\
			& Bias    & -0.6C & -2.2C & -4.0C & -6.3C & -7.9C & -9.3C \\
			\midrule
			Persistence & RMV     & 12.0C & 16.2C & 19.7C & 22.7C & 28.2C & 31.7C \\
			& MAD     &   8.2C & 11.1C & 13.6C & 15.8C & 18.5C & 20.5C \\
			& Bias    & -0.4C &  0.6C &  2.0C &  3.5C &  5.5C &  6.8C \\
			\bottomrule
		\end{tabular}
	\end{minipage}
	\begin{minipage}{.33\textwidth}
		\centering
		{Intensity \revtwo{Verification}}\\
		\begin{tabular}{r|rr}
			\toprule
			Lead time   &   6 h &    12 h \\
			\midrule
			RMSE        & 4.9kt & 9.5kt \\
			MAE         & 3.5kt & 7.1kt \\
			Bias        & 0.6kt & 2.3kt \\
			\bottomrule
		\end{tabular}
	\end{minipage}
	\caption{\textbf{Overall Model \revtwo{Verification} at $\mathbf{n=16}$:} (Left) \revtwo{Trajectory verification of structural forecasts, compared to IR persistence forecasts where the radial profiles are fixed at their 0 h values}. Simulation noise (root mean variance and mean absolute deviation) grows rapidly in the first 6 hours; bias increases in magnitude steadily. We note that persistence offers a less biased IR forecast on average, but higher overall errors in structure at all lead times. (Right) Intensity \revtwo{verification} vs HURDAT2 best-track intensities from 2013-2020 at each lead time.}
	\label{table:overall}
\end{table*}

\begin{table*}[p!]
	\begin{minipage}{.49\textwidth}
		\centering
		{Shear Magnitude}\\
		\begin{tabular}{r|ccr}
			\toprule
			&    Structural &      NHC OFCL & \\
			Shear   &     \multicolumn{2}{c}{12-h RMSE/MAE/Bias} & N\\
			\midrule
			0-10kt  &  9.3/7.3/-0.7 kt &  9.3/6.4/-1.1 kt & 243\\
			10-20kt &   9.7/6.9/1.0 kt &  8.1/5.7/-0.5 kt & 509\\
			20+kt   &   8.6/6.3/2.1 kt &  7.5/5.1/-0.6 kt & 439\\
			\midrule
			Total   &   9.2/6.8/1.1 kt &  8.1/5.6/-0.6 kt & 1,191\\
			\bottomrule
		\end{tabular}   
	\end{minipage}
	\begin{minipage}{.49\textwidth}
		\centering
		{Shear Direction}\\
		\begin{tabular}{r|ccr}
			\toprule
			&     Structural &      NHC OFCL & \\
			Shear Direction &     \multicolumn{2}{c}{12-h RMSE/MAE/Bias} & N\\
			\midrule
			SW              &   8.6/6.1/-0.1 kt &  7.5/5.4/-1.1 kt & 106 \\
			SE              &    9.1/6.9/0.3 kt &  8.5/5.9/-0.5 kt & 440 \\
			NE              &    9.3/6.6/2.2 kt &  7.9/5.4/-0.6 kt & 575 \\
			NW              &  10.6/8.2/-2.0 kt &  8.9/6.1/-0.9 kt &  70 \\
			\midrule
			Total   &   9.2/6.8/1.1 kt &  8.1/5.6/-0.6 kt & 1,191\\
			\bottomrule
		\end{tabular}
	\end{minipage}
	
	\caption{\textbf{Intensity Guidance \revtwo{Verification} Relative to Shear}: Model \revtwo{verification} binned by 200-850-hPa vertical wind shear, reported as RMSE/MAE/Bias. (Left) The performance of the structural forecasting model does not change meaningfully relative to wind shear magnitude, while the NHC official forecast performs better in higher shear environments. The structural forecast has comparable performance to the NHC official forecasts in low-shear environments. (Right) The performance of the structural model does vary with shear direction. Both the NHC forecasts and the structural model produce higher errors for NW shear (6\% of cases).}
	\label{tab:Validation_vs_nhc_shear}
\end{table*}

\begin{table*}
	\begin{minipage}{.5\textwidth}
		\centering
		{TC Intensity}\\
		\begin{tabular}{lccr}
			\toprule
			&      Structural &      NHC OFCL & \\
			Category &     \multicolumn{2}{c}{12-h RMSE/MAE/Bias} & N\\
			\midrule
			Tropical Depression &     5.5/4.4/3.0 kt &  5.9/4.2/-3.1 kt & 112 \\
			Tropical Storm      &     7.2/5.6/2.8 kt &  6.4/4.4/-0.6 kt & 567\\
			Hurricane           &    10.3/7.7/0.6 kt &  9.3/6.7/-0.6 kt & 355\\
			Major Hurricane     &  14.0/10.7/-5.5 kt &  11.8/8.4/0.9 kt & 157\\
			\midrule
			Total   & 9.2/6.8/1.1 kt   &  8.1/5.6/-0.6 kt & 1,208\\
			\bottomrule
		\end{tabular}
	\end{minipage}
	\begin{minipage}{.49\textwidth}
		\centering
		{Intensity Change}\\
		\begin{tabular}{r|ccr}
			\toprule
			&     Structural &       NHC OFCL & \\
			Evolution &     \multicolumn{2}{c}{12-h RMSE/MAE/Bias} & N\\
			\midrule
			Weakening    &  12.1/8.9/7.7 kt &    7.8/5.4/2.9 kt & 263 \\
			Maintenance  &   6.7/5.3/3.3 kt &   6.1/4.3/-0.0 kt & 497 \\
			Intensifying &  9.8/7.3/-5.5 kt &  10.2/7.2/-3.5 kt & 431 \\
			\midrule
			Total   & 9.2/6.8/1.1 kt   &  8.1/5.6/-0.6 kt & 1,208\\
			\bottomrule
		\end{tabular}  
	\end{minipage}
	
	\caption{\textbf{Intensity Guidance \revtwo{Verification} by TC Intensity}: Model \revtwo{verification} split out by intensity and intensity change, reported as RMSE/MAE/Bias. (Left) Both the structural and NHC official forecasts struggle more with intense storms, which are rarer. The structural forecast has much stronger bias, which is expected due to the heavy influence of persistence features in the absence of environmental predictors. (Right) Similarly, both forecasts perform best during maintenance periods (6-hour change $\le5$ kt in magnitude), overestimate during weakening, and underestimate during intensification. The bias is more pronounced in the structural forecast due to the absence of environmental predictors.}
	\label{tab:Validation_vs_nhc_intensity}
\end{table*}

Tables \ref{tab:Validation_vs_nhc_shear} and \ref{tab:Validation_vs_nhc_intensity} assess the performance of our intensity guidance via structural forecasting at 12-hour lead times and compare it to the NHC's official forecast \revtwo{verification} from 2013-2019 due to availability of  \revtwo{verification} data at time of writing; note that this is a subset of the times reported in Table \ref{table:overall}, consisting of cases where both our structural forecasts and NHC official \revtwo{verification} are available. Overall, the RMSE of the structural forecast is about 1.1 kt larger than the NHC official forecast error as computed by RMSE, and structural forecasts produce roughly twice the bias (1.1 vs -0.6 kt).
The structural forecast sees unchanged MSE with increasing 200-850-hPa vertical wind shear; the bias, however, increases with increasing wind shear (Table \ref{tab:Validation_vs_nhc_shear}). This trend is expected, as the model does not include wind shear as a predictor but instead relies on the positive correlation between shear and asymmetry in IR imagery (as captured by radial profiles computed by quadrant). The NHC official forecast error exhibits a similar, if less pronounced, trend in bias with increasing shear. The direction of shear seems more important, with both our model and the NHC official forecast performing most poorly for NW shear (6\% of cases) and best for SW shear (9\% of cases). The NE and SE cases dominate the overall model performance since they comprise the remaining 85\% of the data set. The disparity between different shear magnitudes and directions could be alleviated in a model which utilizes environmental predictors.

Table \ref{tab:Validation_vs_nhc_intensity} demonstrates similar error trends for both official forecasts and our structural forecasts. Errors tend to increase with TC intensity and with rate of intensification or weakening. The structural model produces higher bias for weaker TCs and lower bias for stronger TCs. Similarly, the structural model tends to overestimate intensities during weakening and underestimate them during intensification. The model errors are comparable to NHC official forecast errors during periods of maintenance and intensification (although bias is higher); it is periods of weakening which tend to be poorly modeled by the structural forecast. We suspect that the inclusion of environmental information could improve fidelity in weakening cases; see \revtwo{Section~\ref{sec:future_work} on Future Work Directions} for a discussion of such avenues for model improvement.

\FloatBarrier

\begin{figure*}[t]
    \centering
    \includegraphics{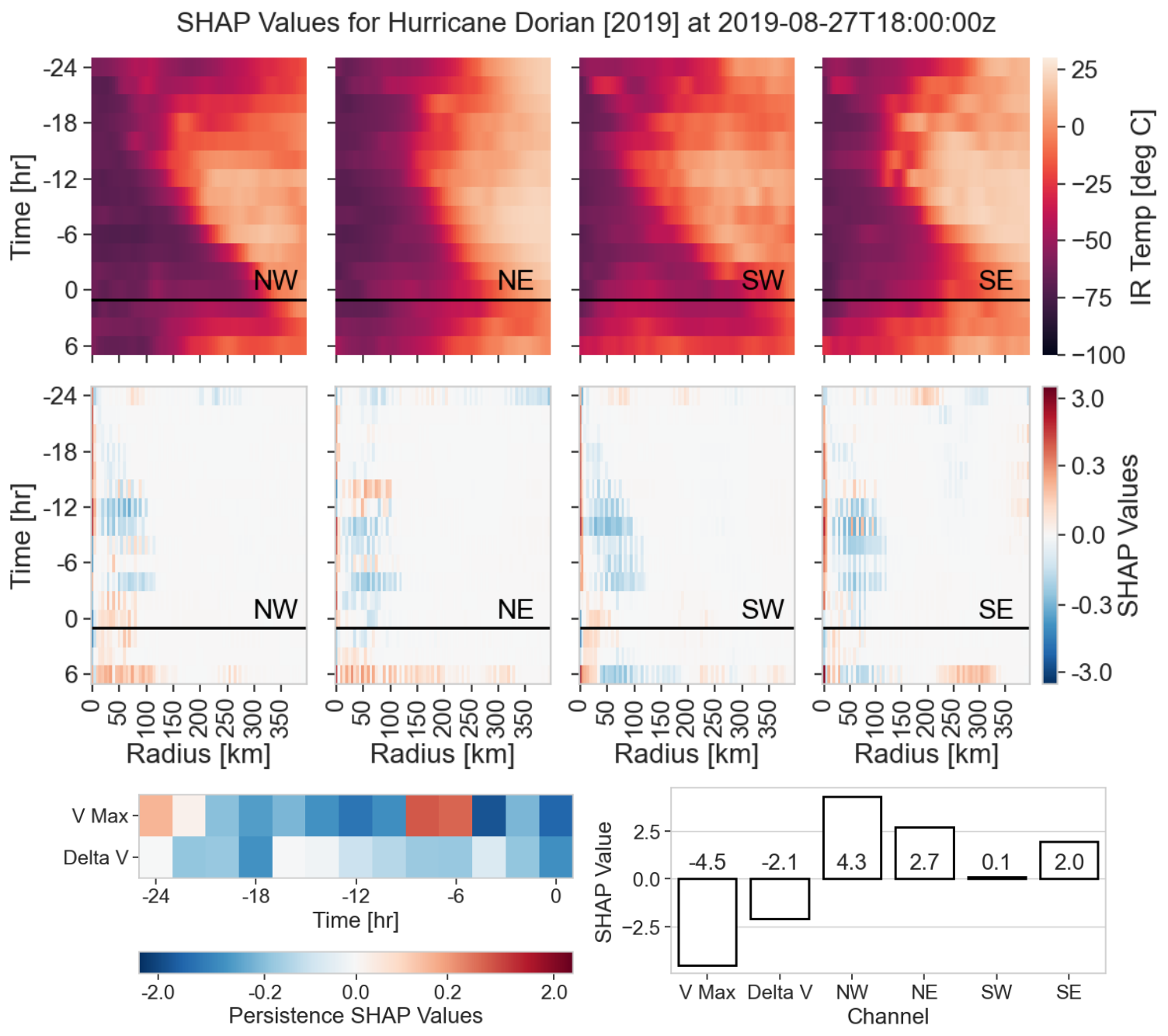}
    \caption{\revtwo{\textbf{SHAP Variable Importance Map for 6-Hour Intensity Forecast:} \revtwo{(Top) IR channels of CNN nowcasting model for Hurricane Dorian [2019] with observed IR structure above the horizontal black lines and +6 h forecasted structure from deep auto-regressive generative model below the horizontal lines. (Center) Pixel-wise SHAP variable importance of IR inputs on the 6-h intensity forecast. (Bottom Left) SHAP variable importance of VMax (linearly interpolated operational intensity estimates) and Delta V (2-h rate of change of operational intensity estimates) on the 6-h intensity forecast.} (Bottom Right) Aggregated SHAP values over each channel, indicating IR features contributing to intensity forecasts to a degree comparable to persistence features.
    }} 
    \label{fig:SHAP_Dorian}
\end{figure*}

\subsection{\revtwo{Variable Importance in Intensity Forecasts}}\label{sec:variable_importance}
\revtwo{Our model results show that structural forecasts result in 6-hour and 12-hour intensity predictions of comparable accuracy to NHC official forecasts. For insight into how much our model relies on IR inputs and prior intensities when making predictions, we compute a saliency map (also known as pixel attribution) for each input. There are varied definitions for saliency, including occlusion-based approaches such as SHAP explainability values \citep{lundberg2017unified}, LIME values \citep{ribeiro2016should}, and gradient-based approaches.}

\revtwo{Figure~\ref{fig:SHAP_Dorian} (center) shows a map of the SHAP importance or contribution of each pixel of the IR observed and forecasted imagery on the 6-hour intensity forecast for Hurricane Dorian [2019]. The bottom-left panel shows the SHAP values for prior intensity and prior intensity change. The bottom-right panel shows aggregated SHAP values for each input channel. From this result and a similar analysis with SHAP variable importance maps for Hurricane Jose [2017] and Hurricane Nicole [2016] in Appendix A and gradient-based saliency maps in Appendix B of Supplementary Materials, we conclude that (i) IR imagery contributes to the intensity forecasts to a degree comparable to persistence features,
%plays as strong or a stronger role in our intensity forecasts than persistence features,
(ii) forecasted infrared imagery from our deep autoregressive generative model plays a more important role than observed past imagery in the TC intensity forecasts, (iii) the current and past presence/absence of an eye is generally the key feature of a storm, and (iv) the core temperatures outside of the eye play a signifcant role for intensity forecasting.}

\section{Discussion and Conclusion}\label{sec:discussion}

This paper demonstrates a novel \revtwo{interpretable} approach to short-term TC intensity guidance \revtwo{trained solely on intensity estimates up to 6 hours prior to the current time and IR observations up to 0 h. We specifically leverage} spatial characteristics of TC convection as captured by radial  \revtwo{IR} profiles}. 
By \revtwo{forecasting an ensemble of +6 h and +12 h trajectories of} TC IR structure  \revtwo{with} radial profiles computed over four geographic quadrants, we obtain reasonable estimates of future \revtwo{+6 h and +12 h} TC intensity while simultaneously capturing \revtwo{and enabling visualization of}  signals in convective structure relevant to those future intensities. We focus on interpretable, physically-based factors to facilitate understanding of the model's performance (e.g., upcoming intensification corresponds with decreasing cloud-top temperatures in the structural forecast). The approach outlined here has the potential for further  \revtwo{improvement by adopting other network architectures for structural forecasts and by}  including environmental predictors provided in real time by SHIPS guidance. Though testing on years of cases takes time, an individual forecast for a single TC can be obtained in minutes on a single GPU, indicating the potential for the eventual use of this model as part of the available TC guidance suite in an operational setting.

\section{\revtwo{Future Work Directions}}\label{sec:future_work}

\subsection{Improving the Network Architecture for Structural Forecasts}
The PixelSNAIL approach provides reasonable simulations of TC IR structural evolution up to 12-hour lead times. However, there exists a wealth of alternate deep autoregressive generative models, each of which can be designed and trained in innumerable ways. Likewise, deep autoregressive models are not the only generative models available. Simulation could be carried out via vector autoregression on a low-dimensional projection of profiles (e.g., principal component analysis, Fourier bases, etc.), generative adversarial networks (GANs; \citealt{creswell2018generative}), \revtwo{or transformers (e.g., temporal fusion transformers for multihorizon forecasting~\citep{lim2021temporal} and spatiotemporal transformers~\citep{grigsby2021long}).} 
The PixelSNAIL architecture  was chosen to demonstrate the value and feasibility of structural forecasting for intensity guidance.

\subsection{\revtwo{Calibrating} the Probability Distribution of Structural Forecasts}
Our structural forecasts are probabilistic in nature, taking the form of {\em probability distributions} over future structural trajectories $\S_{>t}$. In the current work, we apply a standard machine learning approach of fitting a model by minimizing a loss function (in this case the negative log likelihood). A good probabilistic forecast, however, should be {\em conditionally calibrated}. That is, the probability of a particular event (in our case, specific radial profiles 6-12 hours into the future), given or ``conditional on'' a particular history of evolution and other predictors, should match the predicted probability of the same event. This is essentially saying that draws from the forecasting model should be indistinguishable from actual observations, if all relevant conditions are the same. \cite{dey2022calibrated} recently proposed a new method for adjusting or ``recalibrating'' probabilistic forecasts, so that they will have his property. Indeed, one can potentially apply their procedure sequentially to each autoregressive component $p(Z_{iq}|\Z_{i-1},\dots,\Z_1)$, for pixel $i=1,\dots,n$, and quadrant $q=1, 2, 3, 4$, so as to obtain a conditionally calibrated density over structural trajectories $\S_{>t}$ given present and past observations; see Discussion in \cite{dey2022calibrated}.

\subsection{Inclusion of Environmental Variables}
The PixelSNAIL model presented here is a purely autoregressive process; that is, it simulates future structural features using only past IR imagery as an input. The inclusion of environmental variables known to impact TCs such as vertical wind shear, atmospheric moisture, or sea surface temperature may improve the accuracy of the forward simulation of radial profiles, particularly of structural evolution beyond 12 hours. Such factors can be added to the PixelSNAIL architecture as additional input layers via values provided by SHIPS which are not forecasted by the model. These inputs would then serve as ``guiderails'' for simulated structural evolution with potential to better capture the effects of such factors on profile asymmetry. \revtwo{Despite these limitations, our prototype model (which is derived solely from prior and present TC intensity estimates and Geo IR imagery alongside forecasted TC structure using a very simple network architecture) provides reasonable short-term structural and intensity forecasts comparable to NHC forecasts at 6- and 12-h lead times. The inclusion of environmental variables in the nowcasting model is likely to improve its intensity forecasts, which would then be compared to SHIPS forecasts as well as NHC official forecasts, the latter of which are crafted using SHIPS and other guidance.}

%%%%%%%%%%%%%%%%%%%%%%%%%%%%%%%%%%%%%%%%%%%%%%%%%%%%%%%%%%%%%%%%%%%%%
% ACKNOWLEDGMENTS
%%%%%%%%%%%%%%%%%%%%%%%%%%%%%%%%%%%%%%%%%%%%%%%%%%%%%%%%%%%%%%%%%%%%%
\acknowledgments
\revtwo{Part of this research was done as an Independent Study in Spring 2021 while Pavel Khokhlov was a Master in Machine Learning student at Carnegie Mellon University. We are grateful to Microsoft for providing Azure computing resources for this work. The authors would like to thank Katerina Fragkiadaki for a discussion on deep generative networks, and Galen Vincent for many  helpful comments on the research.} This work is supported in part by NSF DMS-2053804, NSF PHY-2020295, and the C3.ai Digital Transformation Institute.

%  Keep acknowledgments (note correct spelling: no ``e'' between the ``g'' and
% ``m'') as brief as possible. In general, acknowledge only direct help in
%  writing or research. Financial support (e.g., grant numbers) for the work done, 
%  for an author, or for the laboratory where the work was performed must be 
%  acknowledged here rather than as footnotes to the title or to an author's name.
%  Contribution numbers (if the work has been published by the author's institution 
%  or organization) should be placed in the acknowledgments rather than as 
%  footnotes to the title or to an author's name.

%%%%%%%%%%%%%%%%%%%%%%%%%%%%%%%%%%%%%%%%%%%%%%%%%%%%%%%%%%%%%%%%%%%%%
% DATA AVAILABILITY STATEMENT
%%%%%%%%%%%%%%%%%%%%%%%%%%%%%%%%%%%%%%%%%%%%%%%%%%%%%%%%%%%%%%%%%%%%%
% 
%
\datastatement
Code to generate Geo IR radial profiles from openly available data can be found at \url{https://github.com/ihmcneely/ORB2sample}, which draws from the MERGIR database openly available from NASA at \url{https://disc.gsfc.nasa.gov/datasets/GPM_MERGIR_1/summary}. The HURDAT2 best track database is openly available from the NHC at \url{https://www.nhc.noaa.gov/data}, while the ATCF operational best tracks (B-deck) are openly available from the National Center for Atmospheric Research at \url{http://hurricanes.ral.ucar.edu/repository/}. Official forecast \revtwo{verification} files are openly available from the NHC at \url{https://www.nhc.noaa.gov/verification/}. Finally, SHIPS developmental data is openly available from the Cooperative Institute for Research in the Atmosphere at \url{https://rammb.cira.colostate.edu/research/tropical_cyclones/ships/developmental_data.asp}.

%  The data availability statement is where authors should describe how the data underlying 
%  the findings within the article can be accessed and reused. Authors should attempt to 
%  provide unrestricted access to all data and materials underlying reported findings. 
%  If data access is restricted, authors must mention this in the statement. See
%  {http://www.ametsoc.org/PubsDataPolicy} for more info.

%%%%%%%%%%%%%%%%%%%%%%%%%%%%%%%%%%%%%%%%%%%%%%%%%%%%%%%%%%%%%%%%%%%%%
% APPENDIXES
%%%%%%%%%%%%%%%%%%%%%%%%%%%%%%%%%%%%%%%%%%%%%%%%%%%%%%%%%%%%%%%%%%%%%
%
%% If only one appendix, use

%% If more than one appendix, use \appendix[<letter>], e.g.,

%\appendix[A] 

%% Appendix title is necessary! For appendix title:

%\appendixtitle{Title of Appendix}

%%% Appendix section numbering (note, skip \section and begin with \subsection)
%
% \subsection{First primary heading}

% \subsubsection{First secondary heading}

% \paragraph{First tertiary heading}

%%%%%%%%%%%%%%%%%%%%%%%%%%%%%%%%%%%%%%%%%%%%%%%%%%%%%%%%%%%%%%%%%%%%%
% REFERENCES
%%%%%%%%%%%%%%%%%%%%%%%%%%%%%%%%%%%%%%%%%%%%%%%%%%%%%%%%%%%%%%%%%%%%%
% Make your BibTeX bibliography by using these commands:
\FloatBarrier
\bibliographystyle{ametsocV6}
\bibliography{references}

\onecolumn
\begin{table}
	\centering
	\large
	\bf{Supplemental Material for ``Structural Forecasting for Short-term Tropical Cyclone Intensity Guidance"}
\end{table}

\appendix[A] 
\appendixtitle{\revtwo{Additional SHAP Variable Importance Maps for TC Intensity Forecasts}}
\begin{figure}[b]
	\centering
	\includegraphics{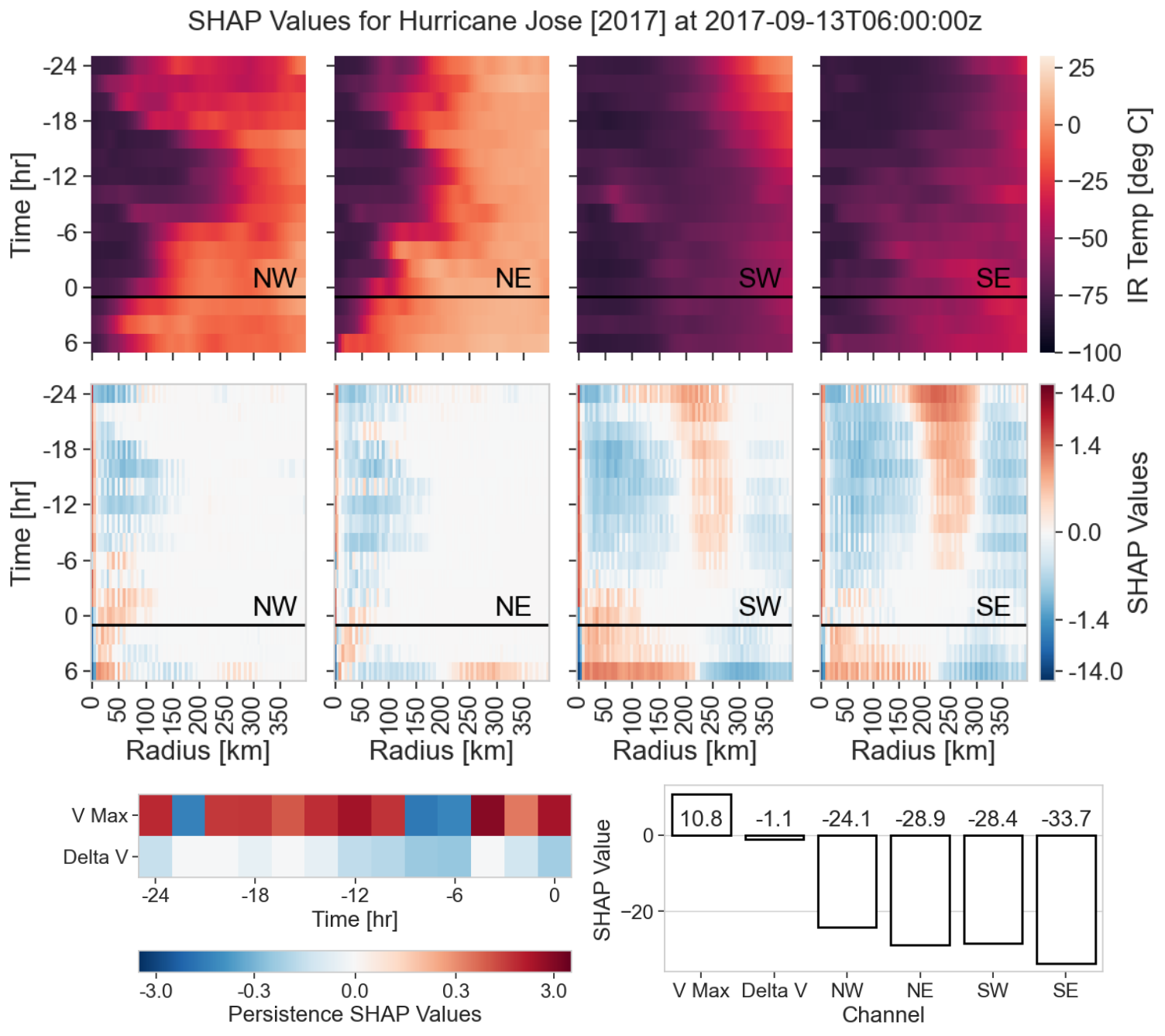}
	\caption{\revtwo{ \textbf{SHAP Variable Importance Map for 6-Hour Intensity Forecast:} (Top) IR channels of CNN ``nowcasting'' model for Hurricane Jose [2017] with observed IR structure above the horizontal black lines, and +6 hour forecasted structure from deep auto-regressive generative model below the horizontal lines. (Center) Pixel-wise SHAP variable importance of IR inputs on 6-hour intensity forecast. (Bottom Left) SHAP variable importance of VMax (linearly interpolated operational intensity estimates) and Delta V (2-h rate of change of operational intensity estimates) on the 6-h intensity forecast.} (Bottom Right) Aggregated SHAP values over each channel showing IR features contributing significantly to the intensity forecasts.
	}
	\label{fig:SHAP_Jose}
\end{figure}
\clearpage
\newpage

\begin{figure}[t]
	\centering
	\includegraphics{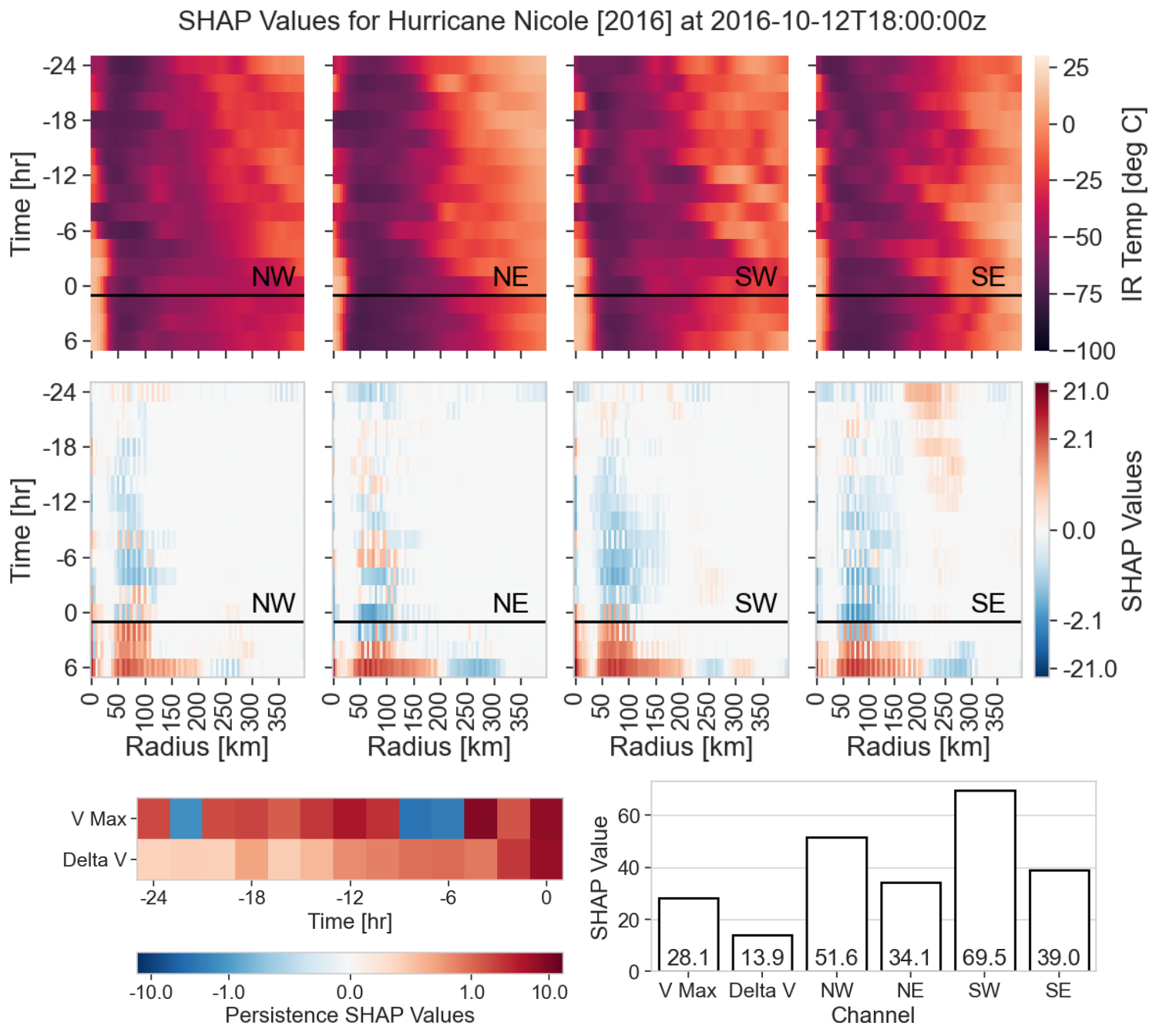}
	\caption{\revtwo{\textbf{IR Channels and SHAP Variable Importance Map for 6-Hour Intensity Forecast:} As figure \ref{fig:SHAP_Jose}, but for Hurricane Nicole [2016].}}
	\label{fig:SHAP_Nicole}
\end{figure}

\clearpage
\newpage

\appendix[B] 

\appendixtitle{Input Saliency for Forecasting Model} 
Our model results in Section 4 showed  that structural forecasts result \revtwo{in 12-hour} intensity predictions of comparable accuracy to NHC \revtwo{official forecasts}. \revtwo{Here we apply a simple gradient-based approach to provide some insight as to how much the model relies on different IR inputs when making predictions. The gradient describes how much a feature contributes to the model response $Y$.} More specifically, we define the saliency $S_i(\x)$ of the $i^{th}$ pixel or feature by
\begin{equation}
S_i(\x)=\Bigg{\lvert}\ \frac{\partial Y}{\partial x_i}\Big\vert_\x\ \Bigg{\rvert},
\end{equation}
\revtwo{}{where $\x$ denotes the total input.}

\begin{figure}
	\centering
	\includegraphics[width=.7\linewidth]{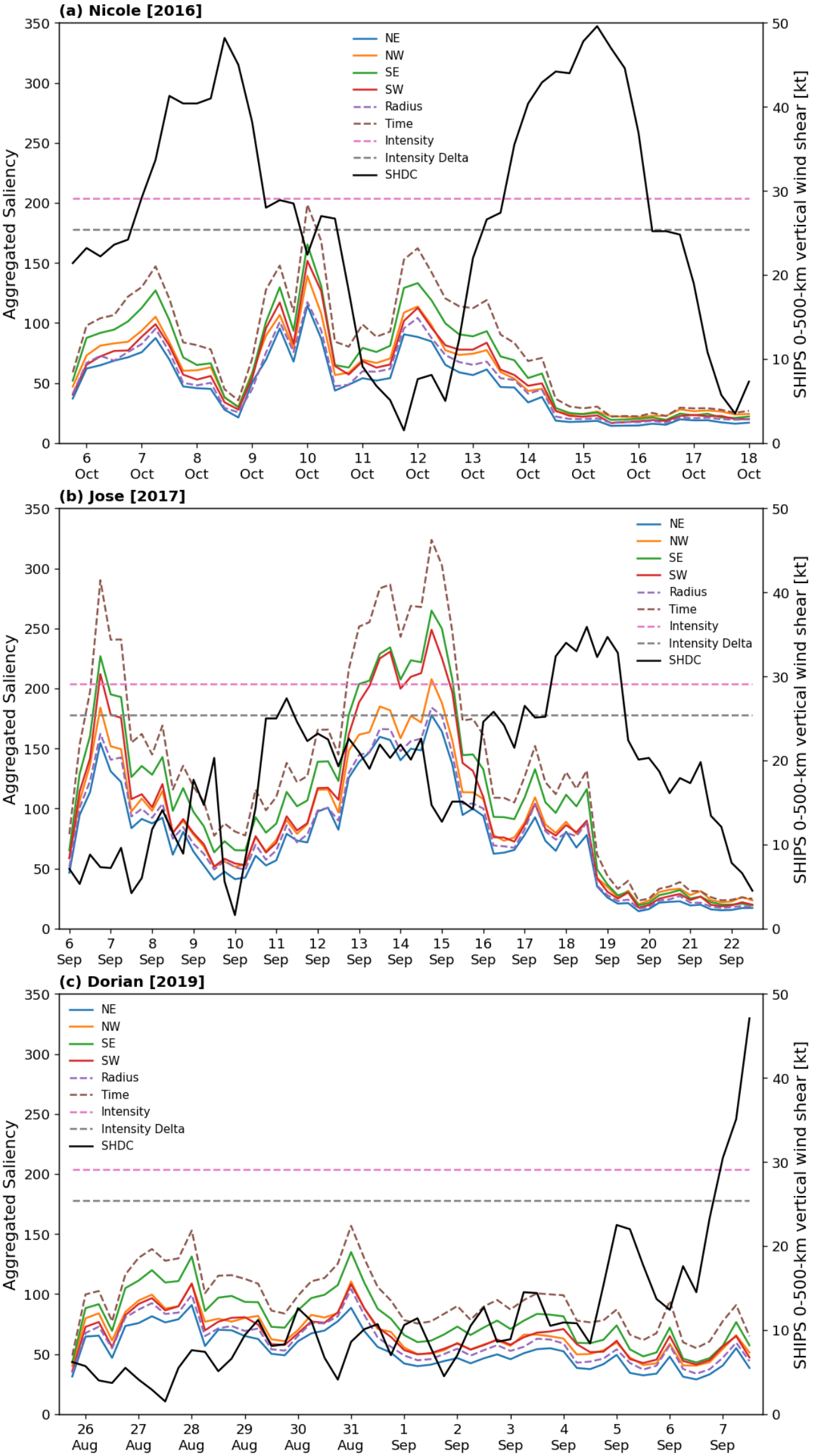}
	\caption{Saliency per input layer in the model. Saliency values over time indicate that the model for TC intensity forecasts utilizes {\em image} inputs to a degree comparable to \revtwo{prior} intensity values.}
	\label{fig:saliency}
\end{figure}

\begin{figure}
	\centering
	\includegraphics[width=.75\linewidth]{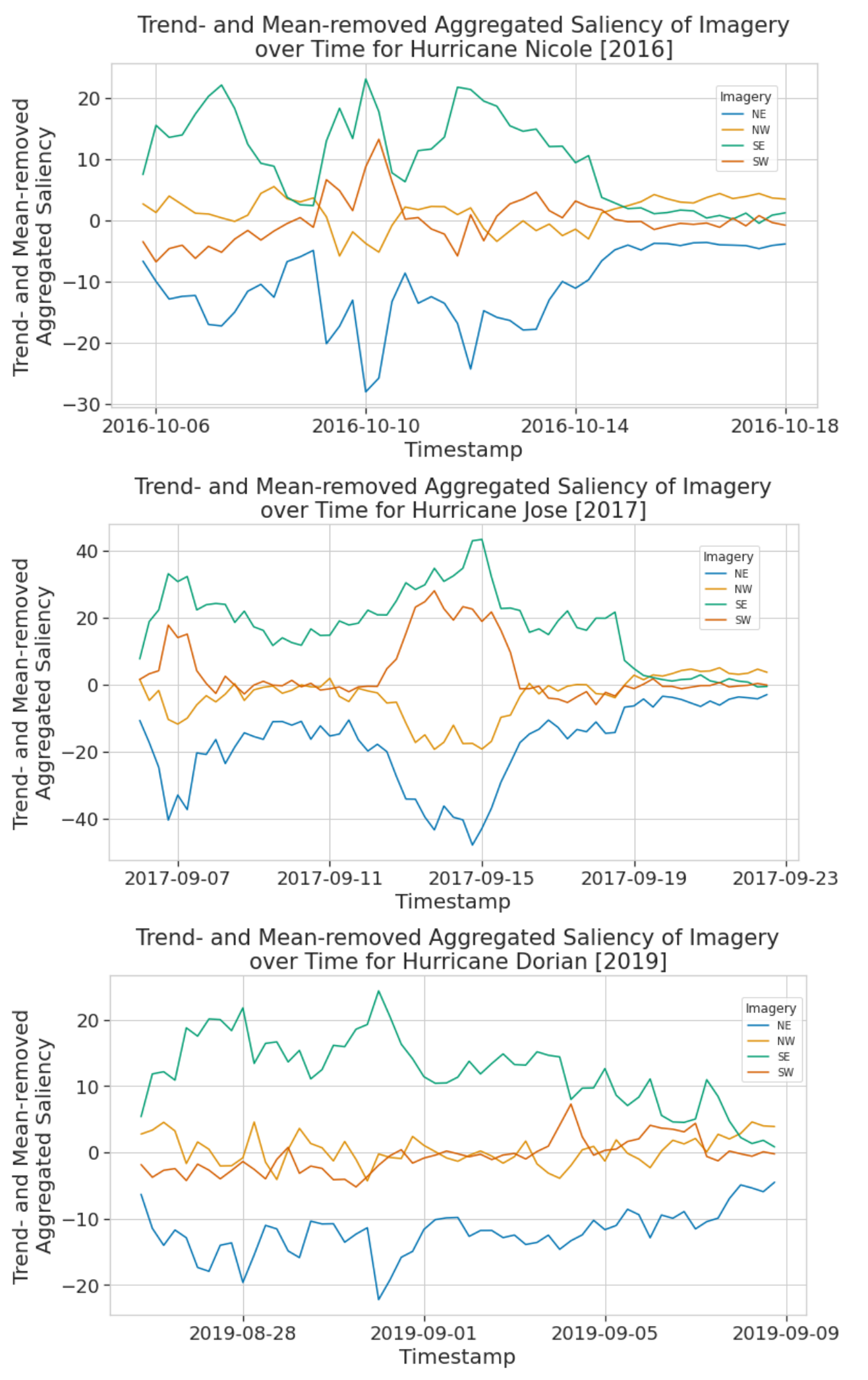}
	\caption{Trend- and mean-removed saliency for image inputs. Comparing the saliency values over time indicates 1) the southern quadrants are more heavily utilized by the forecasting model, and 2) the degree to which the model relies on one quadrant over another is not constant (e.g. Hurricane Jose [2017] around Sept. 14, 2017).}
	\label{fig:saliency_detrend}
\end{figure}

In order to visualize the overall impact of each input channel (four IR quadrants, the radius channel, the time channel, the 
observed \revtwo{prior} intensity, and the 
observed \revtwo{prior} intensity change) on the
\revtwo{forecasted} future intensity, we \revtwo{``aggregate''} the saliency, summing over all pixels in each channel. Figure \ref{fig:saliency} shows the saliency aggregated by channel over time for each of our three example TCs. Note that because 
\revtwo{prior} intensity/intensity change are not included in the convolutional layers, they are linear, and thus have a fixed saliency; because the model is nonlinear in the other channels, the saliency varies over time with the IR inputs. Of particular note is that {\em the aggregated saliency of the IR input channels is comparable to the persistence features}, indicating that the model does not simply rely on persistence to make its predictions but instead makes use of the structural forecasts. Figure \ref{fig:saliency_detrend} shows the same values for IR channels with the mean and trend removed, demonstrating that the model tends to rely more heavily on convective structure in the southern quadrants, and particularly the southeast quadrant.

\clearpage
\newpage

\appendix[C] 
\appendixtitle{Additional Forecast Materials for Case Studies}

\begin{figure}[hp]
	\centering
	\includegraphics[width=\textwidth]{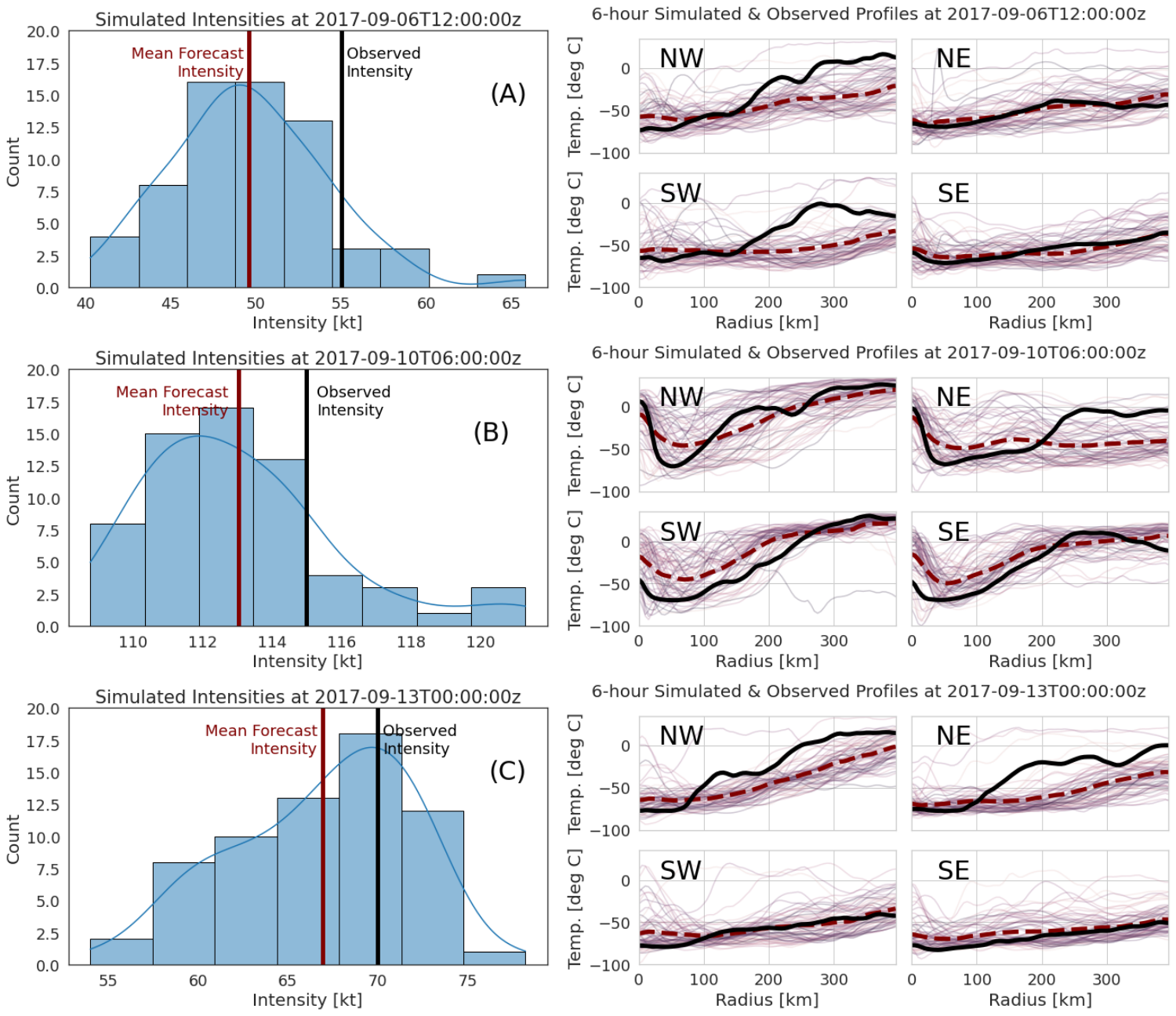}
	\caption{\textbf{Hurricane Jose [2017] 6-hr Guidance:} Hurricane Jose was subjected to vertical wind shear out of the west/northwest due to Hurricane Irma; the structural forecasts tend to underestimate temperatures in the NW quadrant and thus overestimate TC intensity. (Left) Distribution of forecasted intensities with observed (black) and average forecast (red) intensities marked. (Right) Simulated profiles by quadrant with observed profiles represented by solid black curves, and  averaged simulated profiles represented by dashed red curves.}
	\label{fig:Jose_single_time_guidance}
\end{figure}

\begin{figure}[hp]
	\centering
	\includegraphics[width=\textwidth]{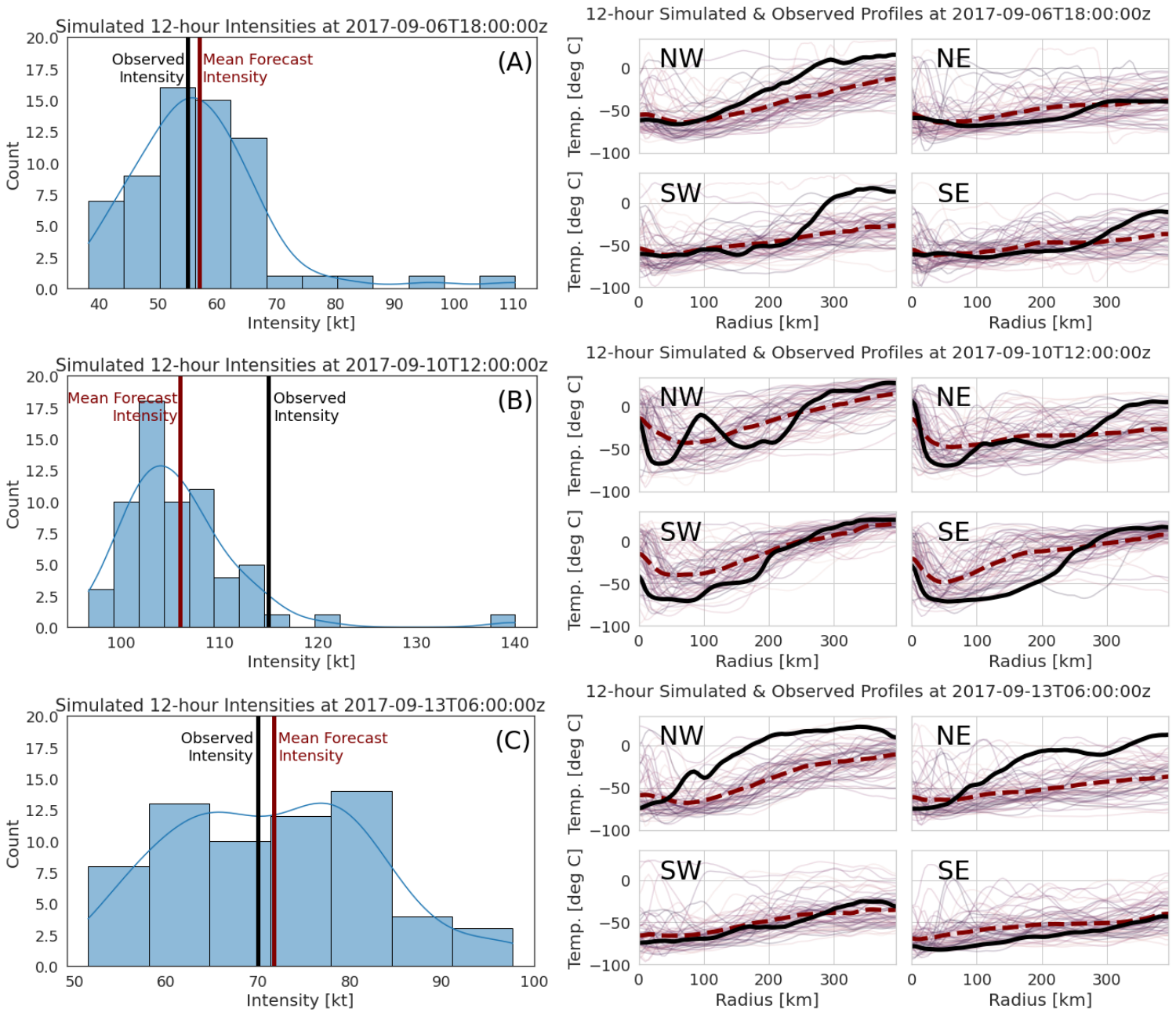}
	\caption{\textbf{Hurricane Jose [2017] 12-hr Guidance:} At 12-hour lead times, this TC's evolution is still well modeled save for a handful of individual overestimates. (Left) Distribution of forecasted intensities with observed (black) and average forecast (red) intensities marked. (Right) Simulated profiles by quadrant with observed profiles represented by solid black curves, and  averaged simulated profiles represented by dashed red curves.}
	\label{fig:Jose_single_time_guidance_12hr}
\end{figure}

\begin{figure}[hp]
	\centering
	\includegraphics[width=\textwidth]{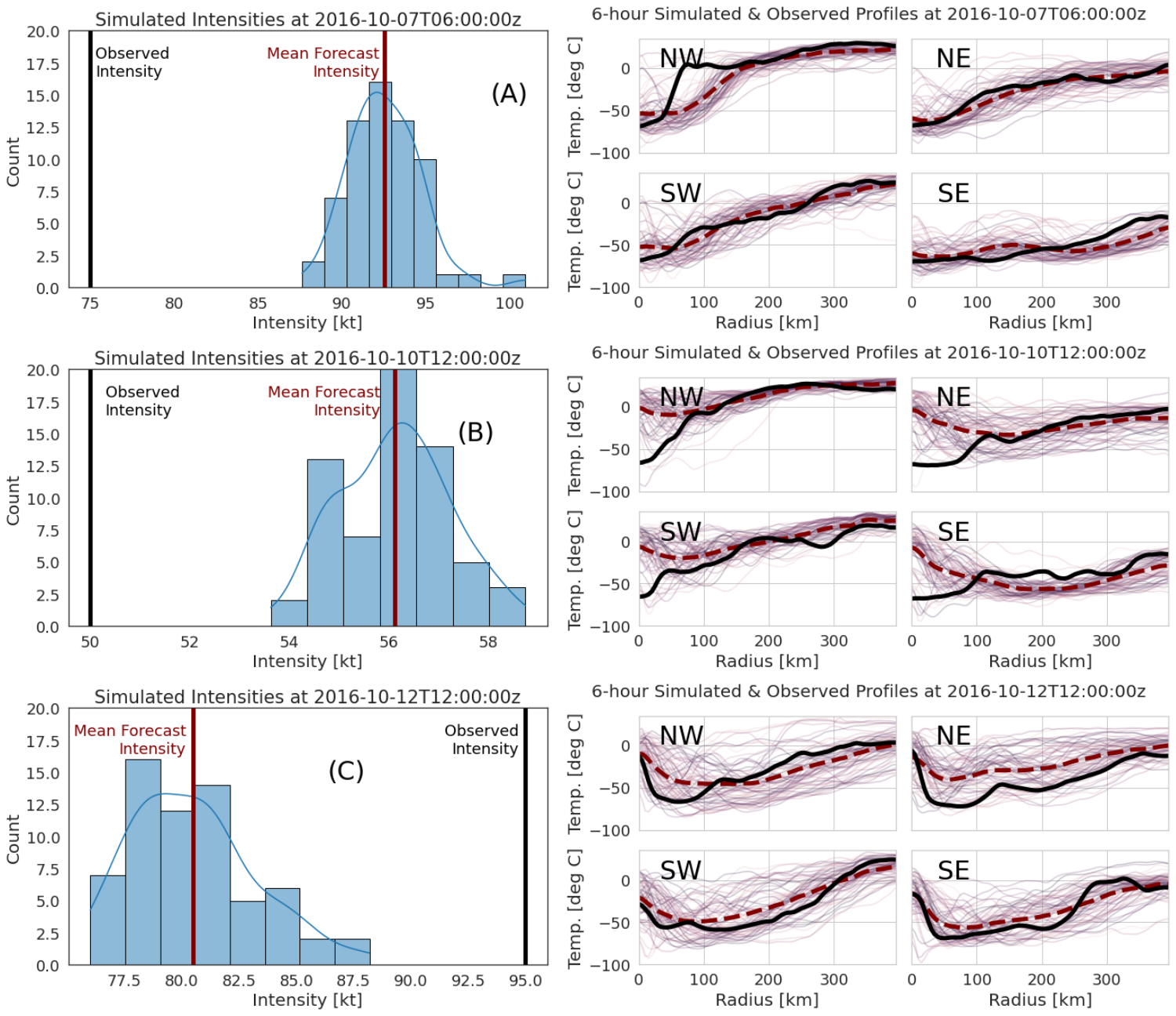}
	\caption{\textbf{Hurricane Nicole [2016] 6-hr Guidance:} Hurricane Nicole underwent several rapid intenity change events. At 6 UTC 7 October, the structural forecast models the cloud top temperatures well, but this is insufficient to predict the extreme change from intensification to weakening even at 6-hour lead times. (Left) Distribution of forecasted intensities with observed (black) and average forecast (red) intensities marked. (Right) Simulated profiles by quadrant with observed profiles represented by solid black curves, and  averaged simulated profiles represented by dashed red curves.}
	\label{fig:Nicole_single_time_guidance}
\end{figure}

\begin{figure}[hp]
	\centering
	\includegraphics[width=\textwidth]{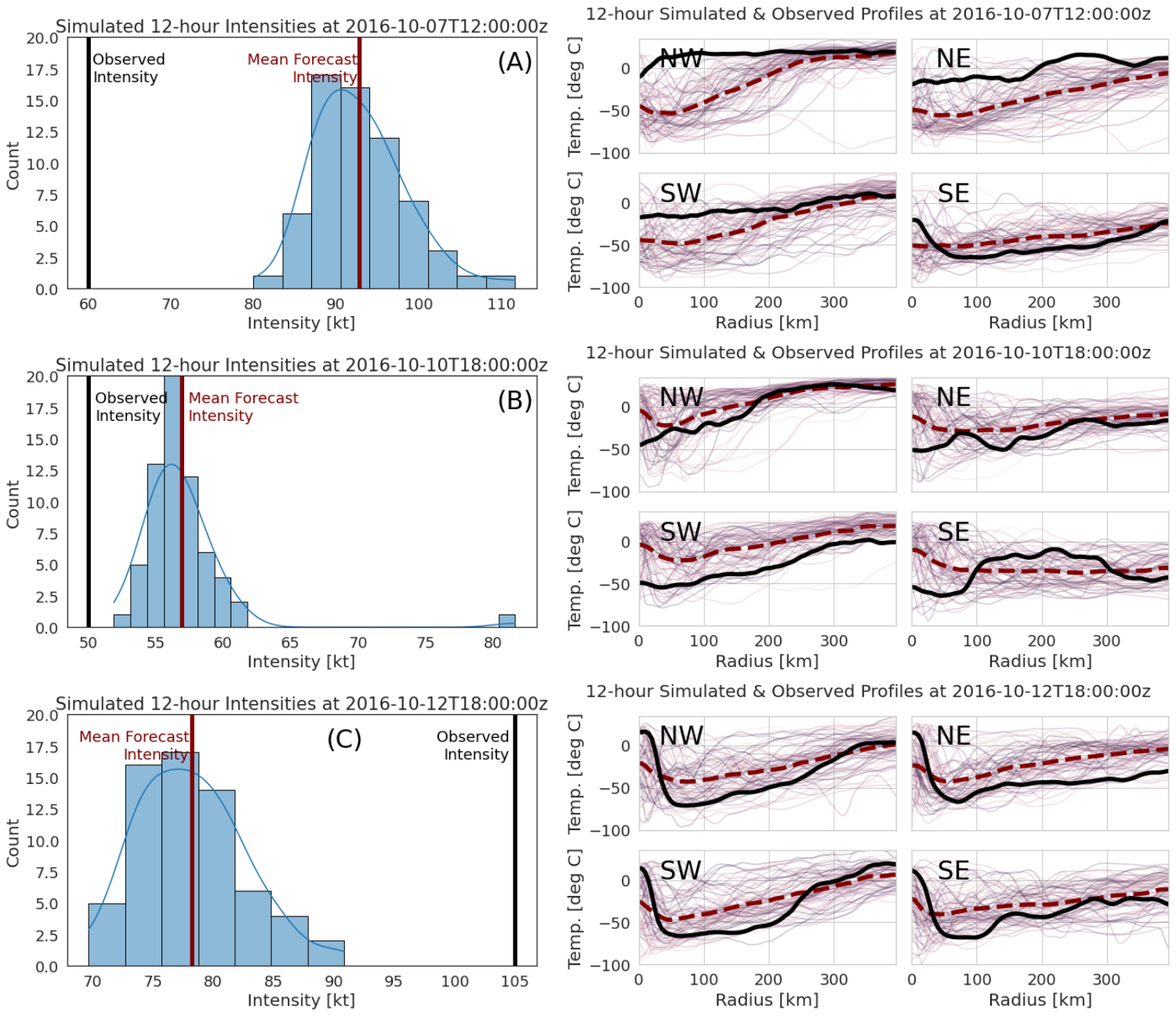}
	\caption{\textbf{Hurricane Nicole [2016] 12-hr Guidance:} Model biases during the quick shifts from intensification to weakening and during stead intensification are exacerbated at 12-hour lead times. (Left) Distribution of forecasted intensities with observed (black) and average forecast (red) intensities marked. (Right) Simulated profiles by quadrant with observed profiles represented by solid black curves, and  averaged simulated profiles represented by dashed red curves.}
	\label{fig:Nicole_single_time_guidance_12hr}
\end{figure}

\begin{figure}
	\centering
	\includegraphics[width=.75\textwidth]{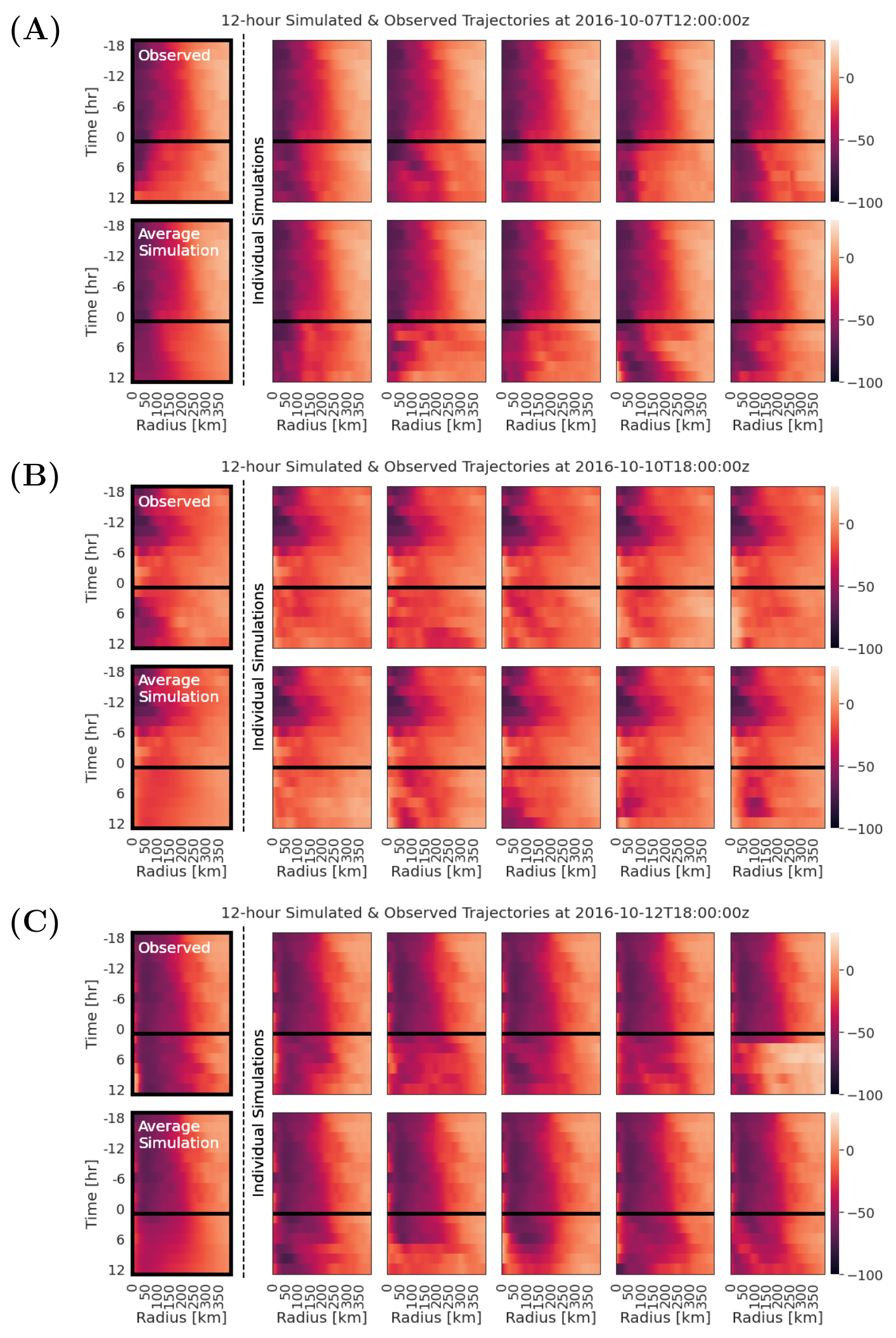}
	\caption{\textbf{Hurricane Nicole [2016] 12-hr Structural Forecasts:} The {\em observed} structural trajectory is shown in the top left corner of each row. To the right, we see 10 {\em individual simulations} of radial profiles (averaged over all quadrants) at 12-hour lead times. All radial profiles above the black horizontal line are observed, while profiles below the black line are simulated. The bottom left corner shows the {\em average simulation} over 64 simulated trajectories.}
	\label{fig:nicole_3sims}
\end{figure}

\begin{figure}
	\centering
	\includegraphics[width=.9\textwidth]{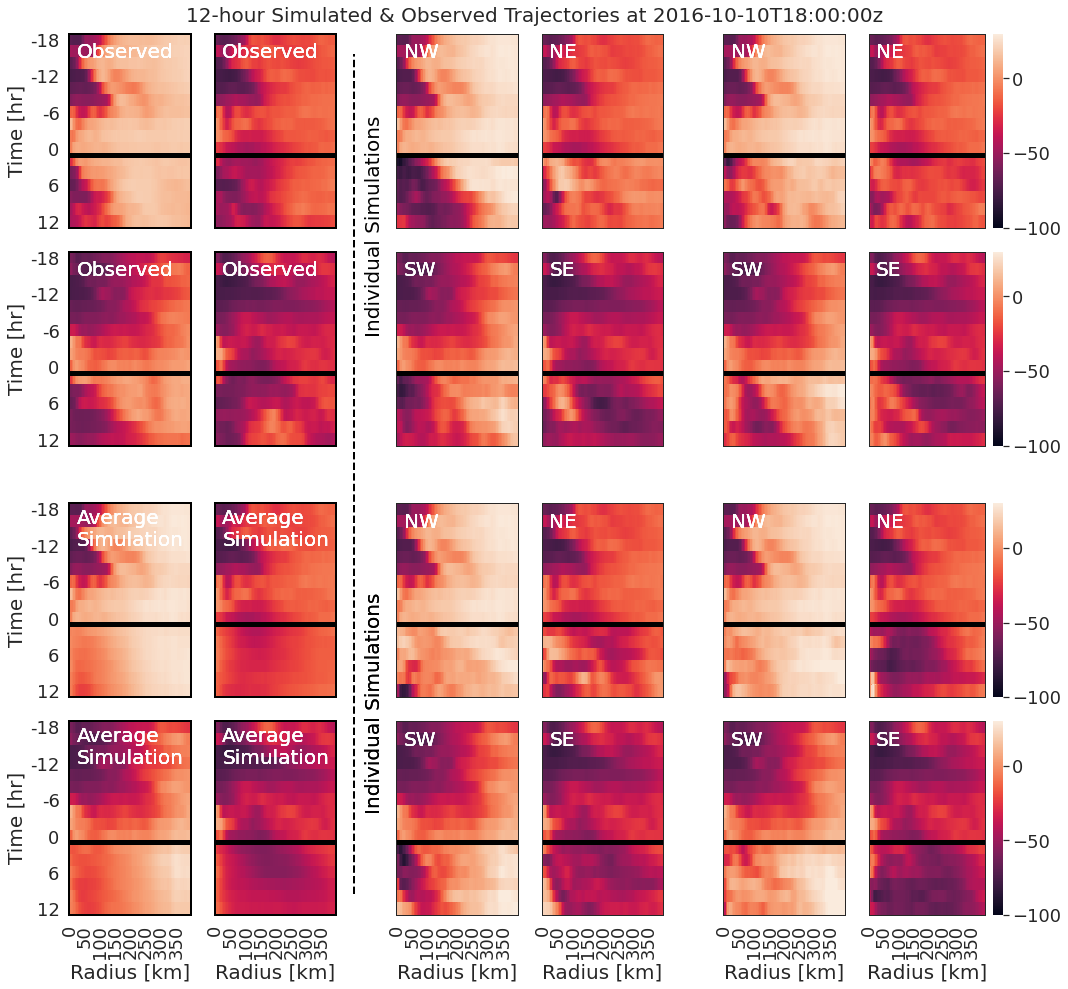}
	\caption{\textbf{Hurricane Nicole [2016] Forecasts By Quadrant (B):} As Figure \ref{fig:nicole_3sims}, but broken down by quadrant for example (B) only. The {\em observed} structural trajectory is shown in the top left corner. To the right, we see four {\em individual simulations} of radial profiles by quadrant at 12-hour lead times. All radial profiles above the black horizontal line are observed, while profiles below the black line are simulated. The bottom left corner shows the {\em average simulation} over 64 simulated trajectories.}
	\label{fig:Nicole_quad_sims}
\end{figure}

\begin{figure}
	\centering
	\includegraphics[width=.75\textwidth]{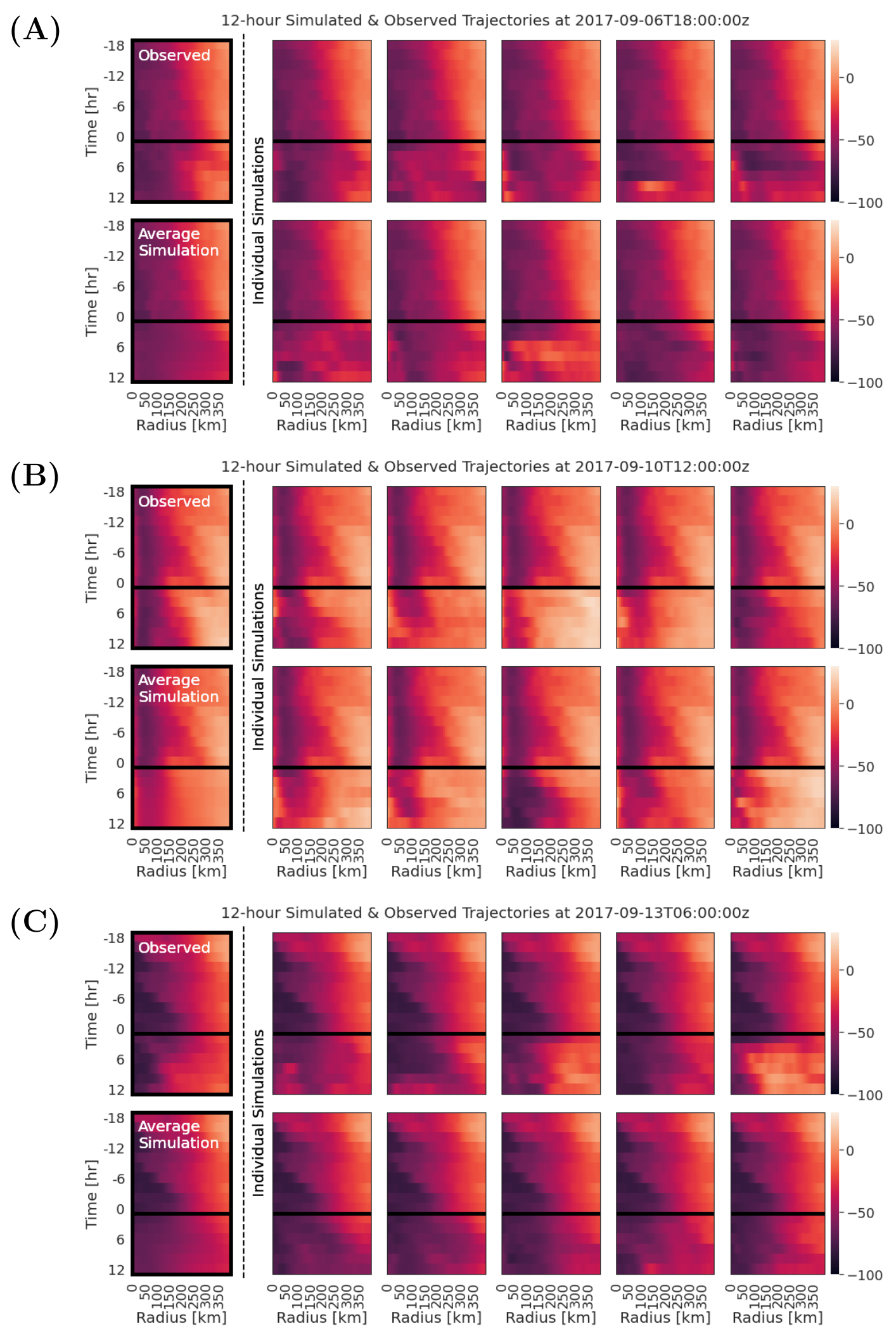}
	\caption{\textbf{Hurricane Jose [2017] 12-hr Structural Forecasts:} The {\em observed} structural trajectory is shown in the top left corner of each row. To the right, we see 10 {\em individual simulations} of radial profiles (averaged over all quadrants) at 12-hour lead times. All radial profiles above the black horizontal line are observed, while profiles below the black line are simulated. The bottom left corner, shows the {\em average simulation} over 64 simulated trajectories.}
	\label{fig:jose_3sims}
\end{figure}

\begin{figure}
	\centering
	\includegraphics[width=.9\textwidth]{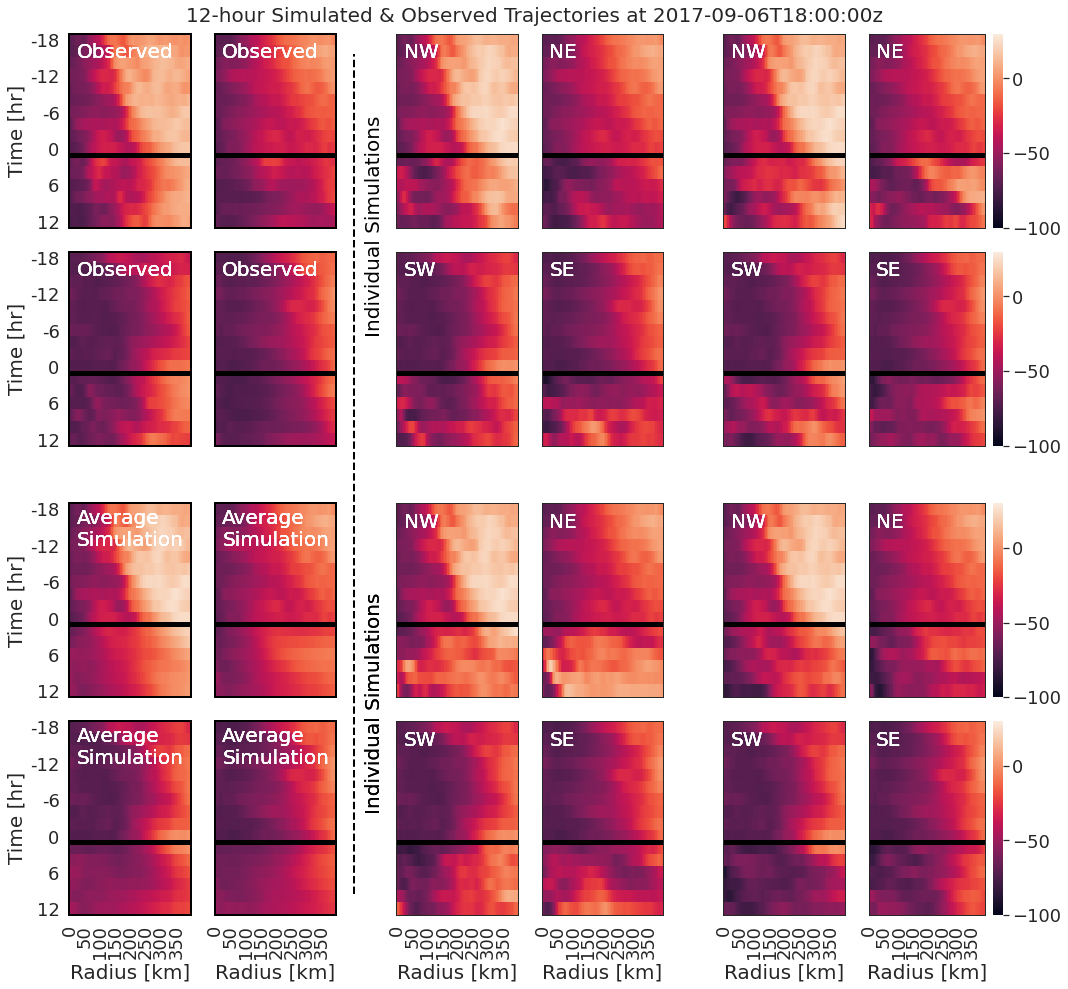}
	\caption{\textbf{Hurricane Jose [2017] Forecasts By Quadrant (A):} As Figure \ref{fig:jose_3sims}, but broken down by quadrant for example (A) only. The {\em observed} structural trajectory is shown in the top left corner. To the right, we see four {\em individual simulations} of radial profiles by quadrant at 12-hour lead times. All radial profiles above the black horizontal line are observed, while profiles below the black line are simulated. The bottom left corner shows the {\em average simulation} over 64 simulated trajectories.}
	\label{fig:Jose_quad_sims}
\end{figure}

\begin{figure}
	\centering
	\includegraphics[width=.75\textwidth]{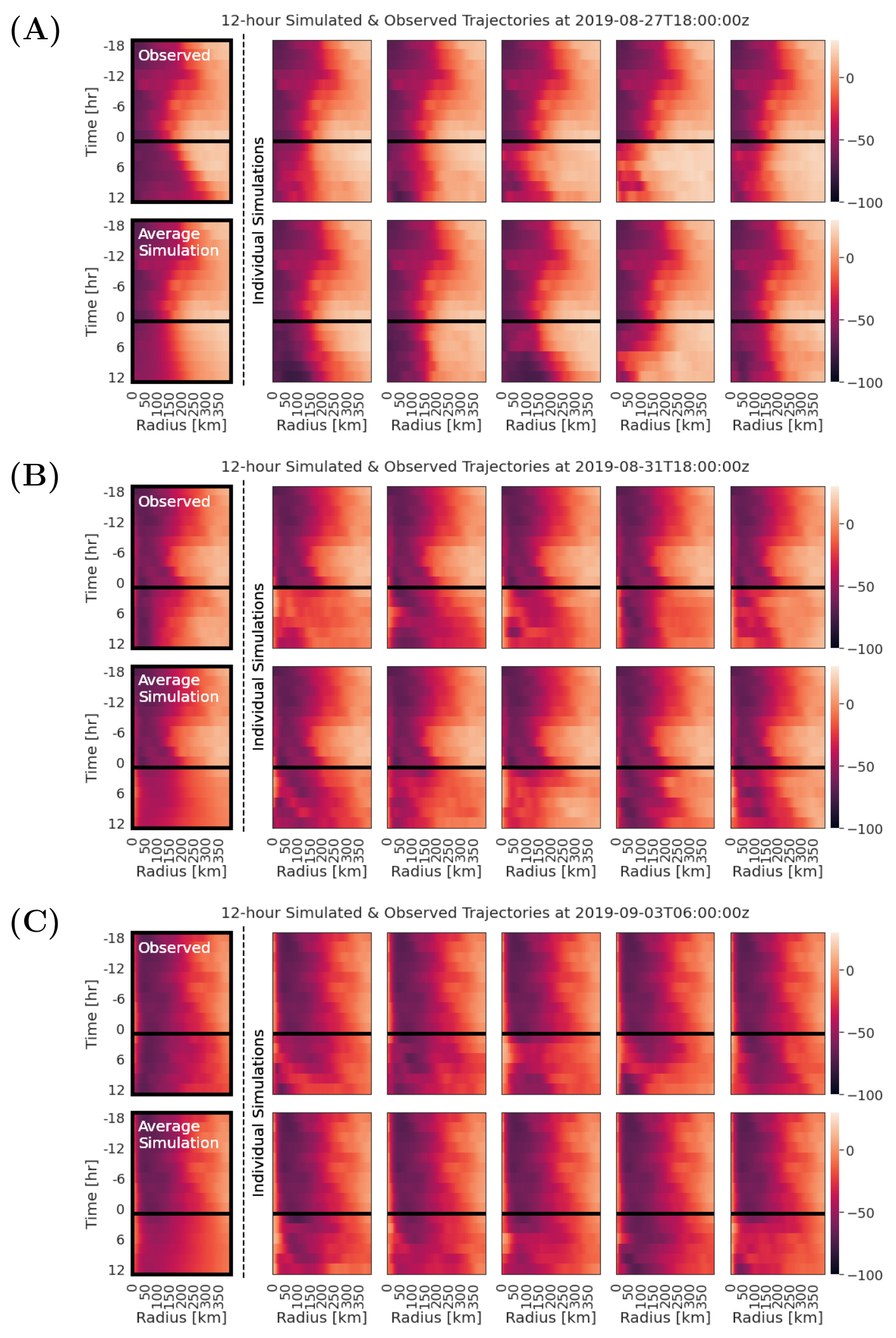}
	\caption{\textbf{Hurricane Dorian [2019] 12-hr Structural Forecasts:} The {\em observed} structural trajectory is shown in the top left corner of each row. To the right, we see 10 {\em individual simulations} of radial profiles (averaged over all quadrants) at 12-hour lead times. All radial profiles above the black horizontal line are observed, while profiles below the black line are simulated. The bottom left corner, shows the {\em average simulation} over 64 simulated trajectories.}
	\label{fig:dorian_3sims}
\end{figure}

\begin{figure}
	\centering
	\includegraphics[width=.9\textwidth]{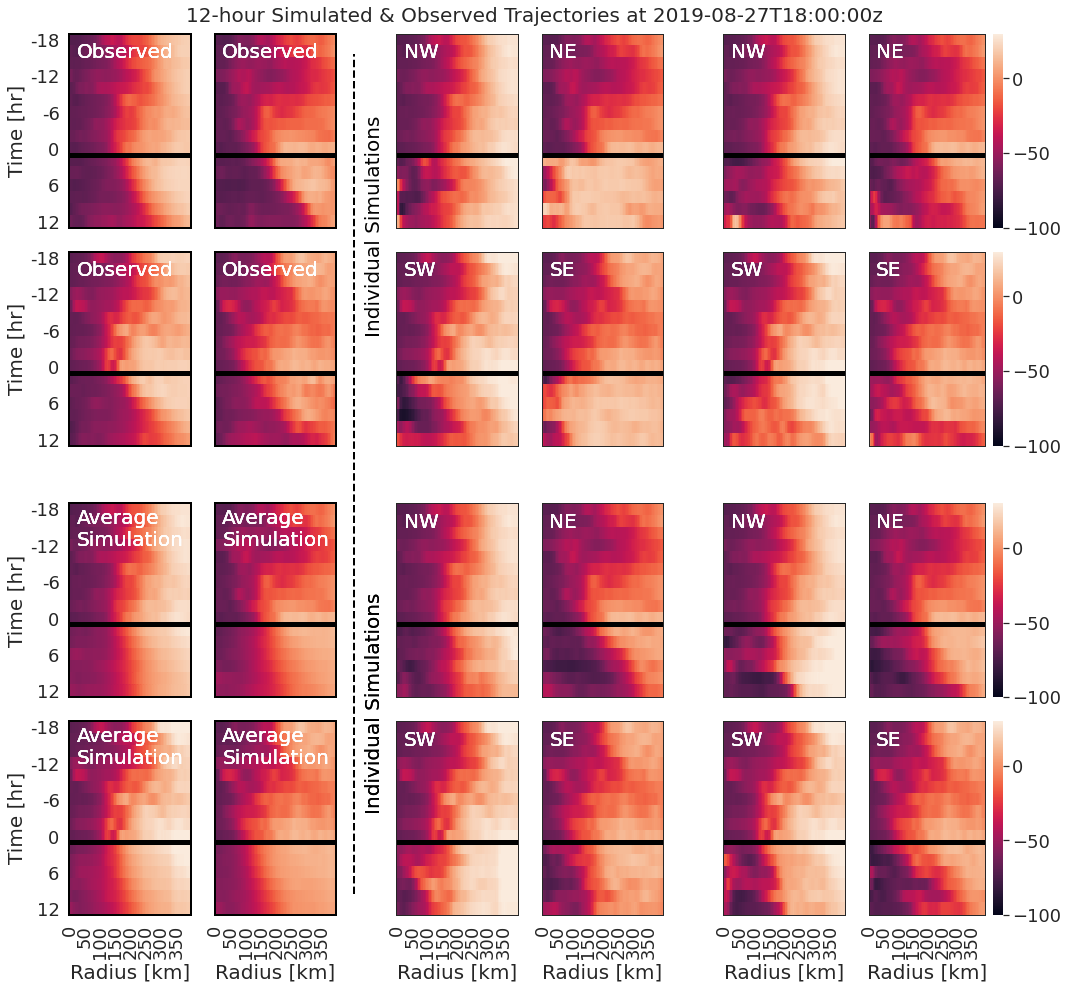}
	\caption{\textbf{Hurricane Dorian [2019] Forecasts By Quadrant (A):} As Figure \ref{fig:dorian_3sims}, but broken down by quadrant for example (A) only. The {\em observed} structural trajectory is shown in the top left corner. To the right, we see four {\em individual simulations} of radial profiles by quadrant at 12-hour lead times. All radial profiles above the black horizontal line are observed, while profiles below the black line are simulated. The bottom left corner shows the {\em average simulation} over 64 simulated trajectories.}
	\label{fig:Dorian_quad_sims}
\end{figure}

\end{document}